\documentclass[12pt,english]{article}
\pdfoutput=1
\usepackage[T1]{fontenc}
\usepackage[utf8]{inputenc}
\usepackage{subfigure,lmodern, amsmath,amssymb, graphicx, pifont, adjustbox, bm, xcolor}
\usepackage{amsfonts}
\usepackage{geometry}
\usepackage{comment}
\usepackage{mathtools}
\usepackage{float}
\usepackage{slashed}
\usepackage{cite}
\usepackage{ragged2e}
\usepackage{etoolbox}
\usepackage{array}
\usepackage{enumitem}

\apptocmd{\thebibliography}{\justifying\setlength{\leftskip}{7.4mm}}{}{}
\makeatletter\g@addto@macro\bfseries{\boldmath}\makeatother

\usepackage{stackengine}
\usepackage{esint}
\usepackage[unicode=true,pdfusetitle,
 bookmarks=true,bookmarksnumbered=false,bookmarksopen=false,
 breaklinks=false,pdfborder={0 0 1},backref=false,colorlinks=true]
 {hyperref}
\hypersetup{pdfauthor={Clifford Cheung and Grant N. Remmen},
 citecolor=black,linkcolor=black,urlcolor=black}

\newcommand{\appendixref}[1]{\hyperref[#1]{appendix~\ref{#1}}}
\def\equationautorefname~#1\null{eq.\,(#1)\null}
\usepackage{breakurl}
\usepackage[hang,flushmargin]{footmisc} 
\allowdisplaybreaks

\usepackage{changepage}

\newcommand{\be}{\begin{equation}}
\newcommand{\ee}{\end{equation}}
\newcommand{\bea}{\begin{eqnarray}}
\newcommand{\eea}{\end{eqnarray}}

\newcommand{\eq}[2]{\be\begin{aligned}#1 \label{#2}\end{aligned}\ee}

\newcommand{\Eq}[1]{Eq.~(\ref{#1})}

\newcommand{\Sec}[1]{Sec.~\ref{#1}}

\newcommand{\dr}{{d}}
\newcommand{\dlog}{d{\rm log}}
\newcommand{\h}{h}
\newcommand{\Res}[1]{\mathop{{\rm Res}}_{#1}}

\newcommand{\AV}{A_V }
\newcommand{\RV}{R_V }
\newcommand{\AH}{A_{\rm hard}}
\newcommand{\AEFT}{A_{\rm EFT}}
\newcommand{\nuP}{\nu^{(P)}}
\newcommand{\nuQ}{\nu^{(Q)}}

\newcolumntype{P}[1]{>{\centering\arraybackslash}p{#1}}

\begin{document}

\interfootnotelinepenalty=10000
\baselineskip=18pt
\hfill CALT-TH 2023-026
\hfill

 \vspace{2cm}
\thispagestyle{empty}
\begin{center}
{\LARGE \bf
Bespoke Dual Resonance
}\\
\bigskip\vspace{1cm}{
{\large Clifford Cheung${}^a$ and Grant N. Remmen${}^{b,c}$}
} \\[7mm]
 {\it ${}^a$Walter Burke Institute for Theoretical Physics, \\[-1mm]
    California Institute of Technology, Pasadena, CA 91125 \\[1.5mm]
${}^b$Kavli Institute for Theoretical Physics and Department of Physics,\\[-1mm] University of California, Santa Barbara, CA 93106 \\[1.5mm]
${}^c$Center for Cosmology and Particle Physics, Department of Physics,\\[-1mm] New York University, New York, NY 10003} \let\thefootnote\relax\footnote{e-mail: \url{clifford.cheung@caltech.edu}, \url{grant.remmen@nyu.edu}} \\
 \end{center}
\bigskip
\centerline{\large\bf Abstract}

\begin{quote} \small

Dual resonance is one of the great miracles of string theory.  At a fundamental level, it implies that the particles exchanged in different channels are subtly equivalent.  Furthermore, it is inextricably linked to the property of exceptionally tame high-energy behavior.  In this paper, we present explicit, closed-form expressions for a new class of dual resonant amplitudes describing an infinite tower of spins for an arbitrary mass spectrum.  In particular, the input of our construction is a user-defined, fully customizable choice of masses.
The resulting ``bespoke'' amplitudes are well behaved in the ultraviolet and analytic except at simple poles whose residues are polynomial in the momentum transfer, in accordance with locality.  The absence of branch cuts can be seen using Newton's identities, but can also be made manifest by expressing the amplitudes as a simple $\dlog$ integral of the Veneziano amplitude that remaps the linear Regge trajectories of the string to a tunable spectrum.   We identify open regions of parameter space that firmly deviate  from string theory but nevertheless comport with partial wave unitarity.  Last but not least, we generalize our construction to the scattering of any number of particles in terms of a $\dlog$ transform of the Koba-Nielsen worldsheet integral formula.

\end{quote}
	
\setcounter{footnote}{0}

\setcounter{tocdepth}{2}
\newpage
\tableofcontents

\newpage

\section{Introduction}

The perturbative scattering amplitudes of string theory are extraordinary mathematical objects.  They seamlessly exhibit an array of miraculous properties that are logically intertwined.  At the heart of these exceptional attributes is the famously tame behavior of string scattering at high energies.  In particular, in the high-energy Regge limit defined by $s\rightarrow\infty$ at {\it some} fixed $t$, perturbative string amplitudes actually vanish,  so 
\eq{
A(s\rightarrow \infty , t) =0.
}{Regge_vanish}    
This behavior is far softer than would be typically expected from scattering amplitudes in a local quantum field theory.

Improved high-energy behavior places strong constraints on the spectrum and interactions of the corresponding theory~\cite{Zohar,CheungRemmen2023}. This is evident from the unsubtracted dispersion relation,
\be
A(s,t) = \frac{1}{2\pi i} \oint\limits_{s'=s} \frac{{d} s'}{s'-s}A(s',t) = \sum_{n=0}^\infty \frac{R(n,t)}{\mu(n) - s}  + A_\infty(t),\label{eq:SDR}
\ee
which recasts the original amplitude as a sum over $s$-channel discontinuities arising from tree-level exchanges, plus a boundary term, 
\be 
A_\infty(t)= \frac{1}{2\pi i}\oint\limits_{s=\infty} \frac{{d}s}{s} A(s,t) . \label{eq:disp}
\ee
Note that we have assumed that $A(s,t)$ is a planar amplitude, and thus only exhibits singularities in the $s$ and $t$ channels.

For any $t$ for which \Eq{Regge_vanish} holds, the boundary term is zero and we can write
\eq{
A(s,t) = \sum_{n=0}^\infty \frac{R(n,t)}{\mu(n) - s} ,
}{dual_res_gen}
where $\mu(n)$ denotes the spectrum of squared masses and $R(n,t)$ is the residue at each level.  The representation of $A(s,t)$ in \Eq{dual_res_gen} implies that the amplitude can be rewritten as a sum over contributions in the $s$ channel alone, even when the scattering process also exhibits singularities in the $t$ channel.  This extraordinary property is known as {\it dual resonance}.  Notably, the very existence of dual resonant scattering amplitudes immediately implies that any fundamental distinction between $s$- and $t$-channel singularities is artificial.

In the case that the $s$ channel describes a finite range of spins, $R(n,t)$ will have bounded degree in $t$. The sum in \Eq{dual_res_gen} then implies that the amplitude is a polynomial in $t$ and thus cannot exhibit $t$-channel singularities.  Conversely, for dynamics with spinning states exchanged in both the $s$ and $t$ channels---e.g., in crossing-symmetric amplitudes---dual resonance requires unboundedly high powers in both variables, corresponding to an infinite tower of spinning particles.\footnote{A trivial exception is a theory of only scalars, which can of course always be written as in Eq.~\eqref{eq:SDR}, with $t$-independent numerators.}

Famously, this trio of extraordinary properties---tame high-energy behavior, dual resonance, and an infinite spin tower---are all hallmarks of string theory. 
The avatar of these coincident miracles is the Veneziano amplitude \cite{Veneziano:1968yb}, which was discovered more than half a century ago in the quest to understand the experimentally measured spectrum of mesons.    Arguably, this work marked the initial conception of string theory. Shortly thereafter, it was understood how dual resonance is a manifestation of the deformability of the string worldsheet to exhibit exchanges purely in the $s$ or $t$ channel alone.

Dual resonance imposes an exceedingly nontrivial constraint on the dynamics.  Somehow, the scattering amplitudes of string theory magically accommodate this property.  But are they the unique mathematical objects that can achieve this feat?
In this paper we revisit this question, only to encounter a vast and unexplored space of new amplitudes that exhibit dual resonance for a {\it tunable spectrum}. 
Importantly, our setup offers the tremendous freedom to fix  the masses of any finite number of states  {\it  to any values of our choice.} Hence, our amplitudes are ``bespoke,'' in the sense that they exhibit a spectrum that is fully customizable via the user-defined input of $\mu(n)$.  Like string amplitudes, our bespoke amplitudes describe an infinite tower of higher-spin states, exhibit softened ultraviolet behavior, and are analytic but for simples poles with local polynomial residues.

The basic mathematical building block of our construction is the Veneziano amplitude,
\eq{
\AV (s,t) = \frac{\Gamma(-s) \Gamma(-t) }{\Gamma(-s-t)}  ,
}{eq:Ven}
defined here modulo unimportant polynomial prefactors in $s$ and $t$, which will not matter much for our analysis.  Clearly, the pole structure of this function can be modified simply by mapping $s$ and $t$ to some alternative, possibly nonlinear, functions of $s$ and $t$.  The key obstacle then becomes how to do this remapping without introducing horribly nonanalytic kinematic structures.  As we will show, this can be achieved using a strikingly compact $\dlog$ transform of the Veneziano amplitude,
\eq{
A(s,t) = \oint \frac{\dlog f(s,\sigma)}{2\pi i} \oint \frac{\dlog f(t,\tau)}{2\pi i}A_V(\sigma,\tau),
}{dlog}
which defines a contour integral over $\sigma$ and $\tau$, with $s$ and $t$ to be treated as constants.  Here $f$ is a polynomial whose zeros, $f(s,\sigma)=f(t, \tau)=0$, define a set of {\it multi-branched} functions, $\sigma$ and $\tau$, which depend on $s$ and $t$, respectively.  In particular, the contour of integration encircles the roots of $f$ in the $\sigma$ and $\tau$ planes.    As we will see later, the zero locus of $f(\mu,\nu)=0$ also gives an implicit definition the spectrum of the theory, $\mu(\nu)$.
For example, in the case of $f(s,\sigma) = \sigma-s$ and $f(t,\tau) = \tau-t$, the contour integral is trivial and we simply obtain the Veneziano amplitude.  Furthemore, in this case the zeros of $f(\mu,\nu) = \nu-\mu$ dictates the corresponding string spectrum, $\mu(\nu)=\nu$.   

On the other hand, when $f$ is a nonlinear function, we discover scattering amplitudes that are far richer in structure.  Indeed,  \Eq{dlog} a priori allows for drastic departures from a stringy spectrum, e.g., even allowing for an asymptotically Kaluza-Klein spectrum.  However, we will see that partial wave unitarity disfavors extreme asymptotic deviations, in accordance with known general results on the asymptotic uniqueness of the Veneziano amplitude \cite{Zohar}.
Importantly, for many choices of $f$, the amplitude in \Eq{dlog} is still relatively well behaved in the high-energy Regge limit.  In these cases, the boundary term $A_\infty(t)$ in \Eq{eq:disp} is well  defined and calculable, and can actually be reabsorbed as polynomial $s$-dependence in the residue, yielding
\eq{
A(s,t) = \sum_{n=0}^\infty \frac{R(n,s,t)}{\mu(n) - s} .
}{}
As advertised, our bespoke amplitudes are dual resonant, in the sense that they can be expressed as an infinite sum over contributions in a single channel.

Unfortunately, we do not yet have an {\it analytic} understanding of the full set of necessary or sufficient conditions for which bespoke amplitudes satisfy partial wave unitarity.  Rather famously, even for standard string amplitudes, a rigorous derivation of unitarity from the point of view of scattering amplitudes has only recently emerged~\cite{Arkani-Hamed:2022gsa}.  Nevertheless, it is straightforward to {\it numerically} study partial wave unitarity in a broad class of examples, where we establish open regions that appear to be fully consistent.

Last but not least, we extend our construction to include any number of external particles using the natural generalization of the $\dlog$ transform in \Eq{dlog} to higher-point scattering amplitudes.   Applying this mapping to the Koba-Nielsen formula, we derive explicit integral representations for bespoke amplitudes at higher-point, which potentially offers a hope for some exotic worldsheet formulation of dual resonant scattering for an arbitrary spectrum.

It is worthwhile to contrast the results of the present paper to prior efforts.  
Notably, the bespoke construction described here goes far beyond the original bootstrap approach of Ref.~\cite{CheungRemmen2023}, which derived new dual resonant amplitudes, but still with the usual linear spectrum of string theory.
Furthermore, our bespoke amplitudes are free from peculiar branch-cut singularities, as found in amplitudes with accumulation-point spectra \cite{Coon,CheungRemmen2022}, and free from nonpolynomial residues, as found in certain product-form amplitudes~\cite{Huang:2022mdb}.

Our methodology stands firmly in the spirit of a bottom-up amplitudes bootstrap.   As such, it is far from clear whether these new dual resonant amplitudes actually correspond to fully consistent physical theories.  From a maximally optimistic perspective, these results suggest the enticing possibility of a vast space of theories that achieve the same miracles as string theory but with a different spectrum.  On the other hand, a more sober viewpoint might simply be that scattering amplitudes constraints are {\it not that restrictive}, and hence it is just not exceedingly difficult to construct functions that satisfy the bootstrap conditions.
The overarching question of which mathematical criteria uniquely pick out string amplitudes from the set of all possible functions remains open.  

\section{Review of the Veneziano Amplitude}

As a warm-up, let us briefly summarize some of the special properties of the Veneziano amplitude, which will be a  central mathematical building block in our construction.

\subsection{Masses and Poles}

As is well known, the gamma functions $\Gamma(-s)$ and $\Gamma(-t)$ in the numerator of the Veneziano amplitude in \Eq{eq:Ven}  encode an infinite sequence of poles at $s=n$ and $t=n$ for nonnegative integers $n$.  Here, the spectrum $\mu(n)=n$ describes the familiar linear Regge trajectory of the string.    
Furthermore, the residue at each level is a polynomial in $t$,
\eq{
&\RV(n,t) = \lim_{s\rightarrow n}(n-s)\AV (s,t) =\frac{1}{n!}\frac{\Gamma(t+n+1)}{ \Gamma(t+1) }.
}{eq:RV}
It is straightforward to expand each residue in powers of $t$, 
\eq{
\RV(n,t) = \frac{1}{n!}\sum_{k=0}^n \begin{bmatrix} n+1 \\ k+1 \end{bmatrix} t^k,
}{eq:Stirling}
where the 
quantity in square brackets is the unsigned Stirling number of the first kind.   The formula in \Eq{eq:Stirling} will be especially useful for our later analysis.

\subsection{Asymptotics and Dual Resonance}

Many of the miraculous properties of the Veneziano amplitude follow directly from its exceptionally soft behavior at high energies.  Taking the Regge limit, defined by $s\rightarrow \infty$ at fixed $t$, we obtain 
\eq{
\lim\limits_{s\rightarrow\infty}\AV (s, t) \sim s^t,
}{VenRegge}
dropping all prefactors.
Thus, for any $t<0$, the amplitude exponentially vanishes in the Regge limit.  Note that the inclusion of any additional polynomial factors of $s$ and $t$ will simply shift the exponent in \Eq{VenRegge} by a constant factor.  In this case, the amplitude will still have vanishing Regge behavior for $t$ less than some finite number.

Vanishing Regge behavior implies that the amplitude satisfies the unsubtracted dispersion relation in \Eq{eq:SDR}, since the boundary term is zero.  Hence, it follows that the Veneziano amplitude can be expressed purely in terms of poles in a single channel,
\eq{
\AV (s,t) &=  \sum_{n=0}^\infty \frac{\RV(n,t)}{n-s} .
}{dual_res}
For the Veneziano amplitude, the boundary integral $A_\infty(t)$ in Eq.~\eqref{eq:disp} vanishes for $t<0$ on account of the Regge behavior defined in \Eq{VenRegge}. 

\section{Dual Resonance for a Kaluza-Klein Spectrum}\label{sec:quad}

Rather unintuitively, we can now use the Veneziano amplitude to engineer a dual resonant scattering amplitude that exhibits a distinctly {\it nonstringy} spectrum.  As a toy model, consider the case of a Kaluza-Klein spectrum, 
\eq{
\mu(\nu) = (\nu+\delta)^2,
}{eq:toyspectrum}
where $\nu$ is a nonnegative integer and $\delta$ is an offset characterizing the mass gap. As we will see, this model is not actually realistic.  However, it will cleanly illustrate the underlying mechanics of our framework.

At first pass, there is an obvious way to deform the singularities of the Veneziano amplitude onto a mass spectrum of our choice: simply apply a nonlinear function to $s$ and $t$ before plugging them into the amplitude.   For example, consider 
\eq{
\AV  (-\delta+\sqrt{s}, -\delta+\sqrt{t}),
}{}
which now includes factors of $\Gamma(-\delta+\sqrt{s})$ and $\Gamma(-\delta+\sqrt{t})$ that exhibit simple poles at $s=(n+\delta)^2$ and $t=(n+\delta)^2$ for nonnegative integer $n$.  While this transformation does the trick, the resulting expression now has dangerous branch-cut singularities on account of the newly introduced square roots.   While branch cuts are not intrinsically inconsistent---indeed they typically signal the appearance of multiparticle thresholds---we will opt for a path that ensures that the amplitude only exhibits simple poles.\footnote{The Coon amplitude~\cite{Coon} is a deformation of the Veneziano amplitude that exhibits a branch cut that terminates at the accumulation point of the spectrum.  Recent work indicates possible pathologies very close to the branch point in these amplitudes~\cite{Jepsen}.  At the same time, amplitudes with asymptotic behavior similar to that of the Coon amplitude have also been constructed in explicit, self-consistent string constructions~\cite{Maldacena-Remmen}.  } One path forward is to simply sum over distinct branches of the square root.  The resulting expression is
 \eq{
A(s,t) = &\phantom{{}+{}} \AV  (-\delta+\sqrt{s}, -\delta+\sqrt{t}) +\AV  (-\delta-\sqrt{s}, -\delta+\sqrt{t})\\
&+\AV  (-\delta+\sqrt{s}, -\delta-\sqrt{t})+\AV  (-\delta-\sqrt{s}, -\delta-\sqrt{t}),
}{sum_sqrt}
which has no branch-cut singularities.
As we will see, this prescription is not a singular accident, but rather a specific instance of a very general procedure for constructing dual resonant amplitudes.

\subsection{Masses and Poles}

At this point, it will be convenient to introduce the {\it spectral curve}, which is a polynomial $f(\mu,\nu)$ in two complex numbers $\mu,\nu$, and which implicitly defines the spectrum.   Physically, we should think of $\mu$ as a  ``kinematic variable,'' of the likes of $s$, $t$, or some mass-squared parameter.  On the other hand, $\nu$ is a ``level variable,'' which we can think of as the label indexing each mass resonance.  In particular, while we will sometimes set $\nu$ to an integer---in which case it is literally the discrete label for the spectrum---it should be thought of as an analytic continuation of this label to the complex numbers. 

For the case of a Kaluza-Klein spectrum, the spectral curve is
\eq{
f(\mu,\nu) = (\nu+\delta)^2 - \mu .
}{spectral_KK}
We can solve for the zero locus of this function, $f(\mu,\nu)=0$, in terms of either $\mu$ or $\nu$.  Solving for the former, we obtain \eq{
\mu(\nu) = (\nu+\delta)^2,
}{}
 which by construction exactly reproduces the Kaluza-Klein spectrum of the theory.  Solving for the latter, on the other hand, we obtain two branches of solutions,
\eq{
\nu_{\pm}(\mu) = -\delta \pm \sqrt{\mu}.
}{KK_roots}
These are precisely the functions summed over in \Eq{sum_sqrt}, which we now recognize as simply
\eq{
A(s,t) =  \AV (\nu_+(s),\nu_+(t)) \,{+}\, \AV (\nu_+(s),\nu_-(t)) \,{+}\, \AV (\nu_-(s),\nu_+(t)) \,{+}\, \AV (\nu_-(s),\nu_-(t)).
}{eq:Aquad}
On account of the spectrum of poles, we will refer to Eq.~\eqref{eq:Aquad} as the Kaluza-Klein amplitude. 

Mathematically speaking, \Eq{eq:Aquad} sums over orbits of the Galois group defined by roots of the spectral curve, so we will refer to it as a {\it Galois sum}.  As we will see, this sum effectively removes all branch cuts, leaving only simple poles. As advertised, \Eq{eq:Aquad} only exhibits singularities at $\nu_{\pm}(s) = n$ and $\nu_{\pm}(t) = n$, which correspond to the Kaluza-Klein resonances at $s =(n+\delta)^2$ and $t=(n+\delta)^2$.

We can re-express \Eq{eq:Aquad} in terms of a contour integral,\footnote{Such a contour integral form is possible only because our planar amplitudes have poles that only depend on $s$ or $t$ individually, but not both~\cite{Cattani}.}
\eq{\!
A(s,t) \,{=}\,  \frac{1}{2\pi i}\left(\oint\! \frac{d\sigma}{\sigma{-}\nu_+(s)}{+}\oint\! \frac{d\sigma}{\sigma{-}\nu_-(s)}\right)\! {\times} \frac{1}{2\pi i}\left(\oint\! \frac{d\tau}{\tau{-}\nu_+(t)}{+}\oint\! \frac{d\tau}{\tau{-}\nu_-(t)}\right)\! {\times} \AV(\sigma,\tau),
}{}
where each contour encircles the simple pole in its corresponding integrand.  Combining terms and using the fact that
\eq{
f(\mu,\nu) = \left( \nu- \nu_+(\mu) \right) \left( \nu- \nu_-(\mu) \right),
}{} 
we obtain precisely the $\dlog$ form in \Eq{dlog},
where the contour of integration is the union of loops encircling all the roots of $f(s,\sigma)$ and $f(t,\tau)$ in the $\sigma$ and $\tau$ planes.

\subsection{Asymptotics and Dual Resonance}

Amusingly, we can apply the Galois sum to each term in the dual resonant representation of the Veneziano amplitude in \Eq{dual_res}, yielding a dual resonant representation for the Kaluza-Klein amplitude, 
\eq{
A(s,t) &= \sum_{n=0}^\infty \bigg(  \frac{1}{n-\nu_+(s)}+\frac{1}{n-\nu_-(s)}\bigg)\bigg( \RV(n,\nu_+(t))+ \RV(n,\nu_-(t))\bigg),
}{eq:Galoisfull}
where the sums each run over the pair of conjugate of roots in \Eq{KK_roots}.
The Galois sum over the $s$-channel pole gives
\eq{
 \frac{1}{n-\nu_+(s)}+\frac{1}{n-\nu_-(s)}
= \frac{1}{\mu(n)-s}\times 2(n+\delta),
}{eq:schannelGalois}
which has the desired simple pole at $s=\mu(n)$, times a constant, $s$-independent Jacobian.

Plugging this expression into \Eq{eq:Galoisfull}, we immediately discover that the Kaluza-Klein amplitude has a dual resonant representation,
\eq{
A(s,t)
&= \sum_{n=0}^\infty \frac{R(n,t)}{\mu(n)-s},
}{eq:Aquaddual}
where the residue, $R(n,t)$, is given by the Jacobian times the Galois sum over the residue of the Veneziano amplitude,
\eq{
R(n,t) &= \lim_{s\rightarrow \mu(n)}(\mu(n)-s)A(s,t)\\  
&= 2(n+\delta)\bigg( \RV(n,\nu_+(t))+ \RV(n,\nu_-(t))\bigg).
}{eq:Rtoy}
By explicit calculation we see that
\be 
\begin{aligned}
R(0,t) &= 4\delta \\
R(1,t) &= 4(1+\delta)(1-\delta) \\
R(2,t) &= 2(2+\delta)[(1-\delta)(2-\delta)+t] \\
R(3,t) &= \frac{2}{3}(3+\delta)(2-\delta)[(1-\delta)(3-\delta)+3t],
\end{aligned}
\ee
and so on.
All branch cuts in $s$ and $t$ have cancelled in each term in the dual resonant sum, and thus also in the final amplitude!

Strictly speaking, the dual resonant representation in \Eq{eq:Aquaddual} is only meaningful if the sum converges.  This convergence is intimately connected with the Regge behavior of the Kaluza-Klein amplitude, which we trivially obtain from the Galois sum over the asymptotic behavior of the Veneziano amplitude, 
\eq{
A_\infty(t) = \lim\limits_{s\rightarrow\infty} A(s,t) 
&\sim \lim\limits_{s\rightarrow\infty}  \nu_+(s)^{\nu_+(t)} +\nu_+(s)^{\nu_-(t)}+\nu_-(s)^{\nu_+(t)}+\nu_-(s)^{\nu_-(t)}\\
&\sim  \lim\limits_{s\rightarrow\infty}  (\sqrt{s})^{-\delta  +\sqrt{t}}+ (\sqrt{s})^{ -\delta - \sqrt{t}},
}{KKUV}
dropping terms that are subleading at large $s$.   We emphasize that in calculating the asymptotic behaviors we have dropped all prefactors that either vanish or do not diverge at large $s$.  Consequently, the relative coefficients sitting in front of each term on the right-hand side should not be considered meaningful.

For physical kinematics, corresponding to $t<0$, the exponents in \Eq{KKUV} are complex.  If $\delta=0$, then the large-$s$ scaling is oscillatory, so the asymptotic behavior of the amplitude is ill defined.  Mechanically, this implies that the boundary term $A_\infty(t)$ in \Eq{eq:SDR} is ambiguous and the dual resonant sum is nonconvergent.  On the other hand, if $\delta > 0$, then the asymptotic behavior is damped, and $A_\infty(t)$ actually vanishes.   Conversely, the case of $\delta < 0$ is pathological because the amplitude diverges.   These behaviors can be verified numerically.

Irrespective of these choices, the Kaluza-Klein spectrum asymptotes to $\mu(\nu) \sim \nu^2$ at large $\nu$, which is of course an exceedingly drastic departure from the linear Regge spectrum of the Veneziano amplitude.  Thus, on general grounds~\cite{Zohar}  we expect that the Kaluza-Klein amplitude is inconsistent with partial wave unitarity, and we refer the reader to App.~\ref{app:unitarity} for details.  For this reason, we will not consider this toy model further.

\section{Dual Resonance for an Arbitrary Spectrum}\label{sec:generalmodel}

It is straightforward to generalize the procedure described above to a {\it customizable} spectrum interpolating through any finite number of points of our choosing.    This approach will yield closed-form expressions for bespoke scattering amplitudes exhibiting dual resonance for an arbitrary spectrum.

\subsection{Spectral Curve}

We begin by defining the spectral curve $f(\mu,\nu)$, which is a polynomial whose zero locus,
\eq{
f(\mu,\nu)=0,
}{eq:spectral}
implicitly defines the spectrum of the theory.
Here $\mu$ should be interpreted as a ``kinematic'' argument, which naturally takes on continuous values and corresponds to different choices of $s$ or $t$, depending on the context.  In constrast, $\nu$ should be interpreted as a ``level'' argument, which at nonnegative integer values labels the discrete spectrum of the theory, but can otherwise be analytically continued to any real value.  

For simplicity, this paper will focus solely on the case in which $f(\mu,\nu)$ is linear in $\mu$, 
\be 
f(\mu,\nu) = P(\nu)-\mu Q(\nu), \label{eq:singlebranch}
\ee
where $P$ and $Q$ are polynomials.  
Note that the Kaluza-Klein amplitude in \Sec{sec:quad} is described by this class of functions.  Of course, \Eq{eq:singlebranch} restricts to a small subset of the vast space of possible choices for the spectral curve.  We  leave more thorough analysis of this landscape for future work.

The zero locus in \Eq{eq:spectral} can be solved for either $\mu$ or $\nu$.   If we opt to solve in terms of the variable $\mu$, then the resulting single-branch solution
defines the mass spectrum,
\eq{
\mu(\nu) = \frac{P(\nu)}{Q(\nu)},
}{muPQ}
provided $\nu$ is set to a nonnegative integer, which is a discrete label for the spectrum at each level.  Obviously, if $P$ and $Q$ are of sufficiently high degree, they can be chosen so that the spectrum interpolates through an arbitrarily large but finite number of masses of our choice.   As we will see later, the level number $\nu$ will control the maximal possible spin exchanged at a given level in our bespoke amplitudes, so the spectral curve can be viewed as a parameterization of a general nonlinear Regge trajectory.
 Note that for more elaborate spectral curves than \Eq{eq:singlebranch}, the spectrum itself can be multi-branched.

What are sensible choices for $P$ and $Q$?  To answer this question, the asymptotic behavior of the spectrum at large level $\nu$ is relevant,
\eq{
\lim\limits_{\nu\rightarrow\infty} \mu(\nu) \sim \nu^{|P|-|Q|},
}{large_nu}
where $|P|$ and $|Q|$ are the polynomial degrees of each function.  Let us consider various choices for these relative degrees.

If $|P|\leq|Q|$, then the spectrum has an accumulation point since \Eq{large_nu} asymptotes to a constant.  Such a spectrum is not a priori inconsistent, and in fact amplitudes exhibiting such structure have been studied previously~\cite{Coon,CheungRemmen2022,Maldacena-Remmen,Jepsen,Geiser:2022exp}.  That said, we will see later that our framework for building dual resonant amplitudes simply does not apply to such spectra.

If instead $|P|>|Q|$, then the spectrum diverges.  As we will soon see, our formulation accommodates any spectrum of this type.  However, recall the analysis of Ref.~\cite{Zohar}, which argued that any theory of higher spins that is free from accumulation points must asymptote to a linear Regge spectrum, under some suitable assumptions.  
For simplicity, we hereafter make the assumption that $|P| - |Q|=1$ to comport with that restriction.  For later convenience, we define the coefficients of these polynomials by
\eq{
P(\nu) = \sum_{k=0}^\h p_k \nu^{\h-k}\qquad \textrm{and} \qquad Q(\nu) = \sum_{k=0}^{\h-1} q_{k+1} \nu^{\h-k-1},
}{eq:PQ}
where $\h= |P| = |Q|+1$ denotes the degree of the spectral curve.  In addition, we also take $P$ and $Q$ to be monic, which means that they are polynomials whose leading powers have unit coefficient, so $p_0=q_1=1$. This choice implies fixes our unit conventions so that the asymptotic Regge slope is $\alpha' = \lim\limits_{\nu\rightarrow\infty} \mu(\nu)/\nu =1$.

\subsection{Galois Meets Veneziano}

If we instead solve for the zeros of the spectral curve in \Eq{eq:spectral} in terms of $\nu$, we obtain multiple solutions, which we denote by $\nu_\alpha(\mu)$ where $\alpha = 0,\ldots, \h-1$. By folding these functions into the arguments of the Veneziano amplitude, we obtain a general expression for our bespoke dual resonant amplitude,
\eq{
A(s,t) &=  \sum_{\alpha,\beta}   \AV (\nu_\alpha(s),\nu_\beta(t)), 
}{eq:M}
where $\alpha$ and $\beta$ each run over all zeros of the spectral curve. We refer to \Eq{eq:M} as a {\it bespoke} dual resonant amplitude, since its spectrum of resonances is customizable. Since the summation in \Eq{eq:M} runs over the orbits of the Galois group defined by the spectral curve, we again call this expression a Galois sum. 

As before, it is possible to rewrite the amplitude in \Eq{eq:M} via the residue theorem,
\eq{
A(s,t) &= \frac{1}{2\pi i}  \oint \sum_{\alpha} \frac{\dr\sigma}{\sigma- \nu_\alpha(s)}  \times \frac{1}{2\pi i}  \oint  \sum_\beta \frac{\dr\tau}{\tau- \nu_\beta(t)}\times \AV(\sigma,\tau),
}{eq:dlogpre}
where the contour of integration is simply the sum of infinitesimal loops encircling the singularity in each denominator.  Using that
 the spectral function factorizes into
\eq{
f(\mu,\nu) &= \prod_{\alpha} (\nu - \nu_\alpha(\mu)),
}{spec_curve_fact}
we can then reconstitute the sum over poles in \Eq{eq:dlogpre} into a $\dlog$ integration measure, yielding the  compact expression in \Eq{dlog}.

The bespoke dual resonant amplitude in \Eq{eq:M} has factors of $\Gamma(-\nu_\alpha(s))$ and  $\Gamma(-\nu_\beta(t))$, which for some $\alpha$ and $\beta$ will exhibit singularities when $\nu_\alpha(s)= n$ and $\nu_\beta(t)=n$ for nonnegative integer $n$.  This is guaranteed to occur for some $\alpha$ and $\beta$ since the vanishing of the spectral curve implies that at least one factor of \Eq{spec_curve_fact} vanishes.

The bespoke dual resonant amplitude in \Eq{eq:M} is nothing more than a sum of Veneziano amplitudes evaluated at transformed kinematics.  Thus, one might naively expect that \Eq{eq:M} automatically has the following dual resonant form, 
\eq{
A(s,t) &= \sum_{n=0}^\infty \bigg(  \sum_{\alpha}  \frac{1}{n-\nu_\alpha(s)}\bigg)\bigg( \sum_{\beta} \RV(n,\nu_\beta(t))\bigg),
}{eq:Galoisfull_again}
which is simply the corresponding sum of the dual resonant representation of the Veneziano amplitude.  However, the caveat in this logic is that the above sum need not converge.  More concretely, the unsubtracted dispersion relation in \Eq{eq:SDR} is only well defined if the sum, together with the boundary term $A_\infty(t)$, is well defined.  Since the bespoke dual resonant amplitude is constructed from Veneziano amplitudes evaluated at modified kinematics, the convergence of these boundary terms is not guaranteed and must be checked explicitly.  We will now examine this issue in detail.

\subsection{Asymptotics}\label{sec:control}

It will be illuminating to study our bespoke amplitudes further in various kinematic limits, specifically the Regge, high-energy fixed-angle, and low-energy effective field theory regimes.  As we will see, the corresponding behaviors can differ from those of the Veneziano amplitude.

\subsubsection{Regge Limit}

Let us start by computing the Regge limit, which corresponds to $s\rightarrow \infty$ at fixed $t$.
The Regge behavior is controlled by the zeros of the spectral curve, $f(s,\sigma)=P(\sigma)-s Q(\sigma) = 0$, as $s$ goes to infinity.  Since $|P|=\h$ and $|Q|=\h-1$, the spectral curve is an $h$-degree polynomial in $\sigma$, and thus its zero locus defines precisely $\h$ roots.  Using our earlier notation, we define the roots of $f(s,\sigma)=0$ by the set of functions $\sigma = \nu_\alpha(s)$, which are $s$-{\it dependent} on account of the explicit factor of $s$ in the spectral curve.    
For later convenience, we also define the roots of the separate curves $P(\sigma)=0$ and $Q(\sigma)=0$ by the functions $\sigma = \nuP_\alpha$ and $\sigma = \nuQ_\alpha$, respectively, which are $s$-{\it independent} numbers.  

At large $s$, exactly one of the roots of the spectral curve asymptotes to $s$ itself.  Without loss of generality, we label this branch by $\alpha=0$.  Meanwhile, the remaining $\h-1$ roots of the spectral curve asymptote to the roots of $Q$ and are thus $s$-independent, so 
\eq{
\lim\limits_{s\rightarrow\infty} \nu_\alpha(s) = \left\{ \begin{array}{ll}
s\,,& \quad \alpha=0 \\
\nuQ_\alpha\,,& \quad \alpha=1,\ldots, h-1. \\
\end{array}\right.
}{nulimit}
To determine the Regge behavior of the bespoke dual resonant amplitude, we simply compute the Galois sum of the Regge behavior of the Veneziano amplitude,  
\eq{
A_\infty(t) = \lim\limits_{s\rightarrow\infty} A(s,t) \sim \lim\limits_{s\rightarrow\infty} \sum_{\alpha,\beta} \nu_\alpha(s)^{\nu_\beta(t)} 
\sim \lim\limits_{s\rightarrow\infty} \sum_{\beta} \bigg( s^{\nu_\beta(t)} + \sum_{\alpha\neq0}
\left( \nu_\alpha^{(Q)}\right)^{\nu_\beta(t)}\bigg),
}{asymptotics}
where we have split the sum according to whether a given root diverges with $s$ or not.  Again we emphasize that the relative coefficients between terms on the right-hand side should be ignored, since we have implicitly dropped the prefactors in front of each term that do not diverge at large $s$.

Now we will impose the condition that \Eq{asymptotics} is convergent for some value of $t$.  This is required in order for the unsubtracted dispersion relation in \Eq{eq:SDR}, and thus the dual resonant form of the amplitude, to mathematically exist.  Since the second term in the final expression of \Eq{asymptotics} only involves $s$-independent roots, it does not diverge in the Regge limit, so we need not worry about it.  On the other hand, the first term is manifestly $s$-dependent.  Consequently, in order to ensure convergence for some $t$, we require that
\eq{
{\rm Re}\left(\nu_\beta(t)\right) < 0 \quad \textrm{for all} \quad \beta,
}{Re_cond}
so that the exponent of each of these contributions has negative real component.   

Assuming \Eq{Re_cond}, the first term in the final expression in \Eq{asymptotics} is zero.  Furthermore, we recognize the second term as simply a partial Galois sum over the Regge behavior of the Veneziano amplitude.   We can therefore write the boundary term in \Eq{asymptotics}  as an explicit closed-form expression in terms of $t$,
\eq{
A_\infty(t)=  \sum_{\alpha\neq 0} \sum_\beta A_V( \nuQ_\alpha,\nu_\beta(t)).
}{eq:Galasymp}
Note that the right-hand side is manifestly independent of $s$.
As we will see later, we can use the above formula to check the unsubtracted dispersion relation in \Eq{eq:SDR}.

We have only stipulated the very conservative condition that \Eq{Re_cond} applies for {\it some} value or range of $t$.  Only for this value of $t$ will the boundary term be well defined and will the dual resonant representation of the amplitude exist.  This is not dissimilar from the case in string theory, where a dual resonant representation requires that $t$ be sufficiently negative.

More generally, the negative real condition in \Eq{Re_cond} is a central concept in control theory, and in mathematical parlance is known as {\it Hurwitz stability} of $f(t,\nu)$ as a  polynomial in $\nu$.
Imposing this condition for all $t<0$ is equivalent to requiring stability of all convex combinations of $P$ and $Q$ and is related to Kharitonov's theorem~\cite{Kharitonov,Kale}.
A useful set of necessary and sufficient conditions in terms of the Hurwitz matrix is given in Ref.~\cite{Bialas}.\footnote{The threshold condition on each root $\nu_\alpha(t)$ of $f(t,\nu)=0$ can similarly be shifted by multiplying the amplitude by a crossing-symmetric polynomial in $s$ and $t$ before Galois summing, resulting in a weaker stability condition~\cite{Bialas2}. In any case, one can first characterize the space of stable spectral functions $f$ via the canonical stability problem, first imposing ${\rm Re}(\nu)<0$ for $t<0$ and then shifting $t$ or $\nu$, or both, after the fact. 
Then the generalized version of \Eq{Re_cond} is that there exists {\it some} value of $t$ for which all ${\rm Re}(\nu)$ are upper bounded by {\it some} constant.}

\subsubsection{Hard Scattering Limit}

Next, we compute the hard scattering limit of the amplitude, defined by the function
\eq{
\AH(s,t) &= \! \lim_{s,t\rightarrow\infty}\!  A(s,t)\\
&= \! \lim_{s,t\rightarrow\infty}\!\AV(s,t) \! +\! \sum_{\alpha\neq 0} \!\left(\lim_{s\rightarrow\infty} \!\AV(s,\nuQ_\alpha) \!+ \!\lim_{t\rightarrow\infty}\!\AV(\nuQ_\alpha,t)\right) \! + \!\!\!\sum_{\alpha,\beta\neq 0} \!\!\! \AV(\nuQ_\alpha,\nuQ_\beta),
}{eq:Ahard}
where we have used Eq.~\eqref{nulimit} in order to take the limits of $\nu_\alpha(s)$ and $\nu_\beta(t)$.  Here it will be important to consider separately the case of negative and positive $t$, which exhibit different asymptotic behaviors.

For $t>0$, corresponding to unphysical kinematics, the hard scattering limit of the Veneziano amplitude takes the form
\be 
A_V(s,t) \sim e^{B(s,t)},\qquad B(s,t) = (s+t)\log(s+t)-s\log s-t\log t+\cdots,\label{eq:AVhard}
\ee
which controls the scaling of the first term in Eq.~\eqref{eq:Ahard}.
The second and third terms in Eq.~\eqref{eq:Ahard} are effectively Regge limits, with $s$ taken large and $t$ fixed, or vice versa.
These contributions scale as powers of $s$ and $t$, which are subdominant to the leading contributions in Eq.~\eqref{eq:AVhard}.
Since the fourth term in Eq.~\eqref{eq:Ahard} is a constant, it is also subdominant.
We therefore learn that the hard scattering limit of our bespoke amplitude is the same as for the Veneziano amplitude,
\be 
\AH(s,t) \sim e^{B(s,t)}\qquad \textrm{for} \qquad t>0,
\ee
which is accordance with the general arguments of Ref.~\cite{Zohar}.

For $t<0$, hard scattering corresponds to a physical high-energy process occurring at a fixed scattering angle $\theta$ that satisfies $\cos\theta = 1 + \frac{2t}{s-m_{\rm ext}^2}$, where $m_{\rm ext}^2$ is the mass squared of each external.
In this kinematic region, the leading scaling of $B(s,t)$ is negative, so the Veneziano amplitude exhibits its trademark exponential softness.
Suppose now that the amplitude obeys dual resonance for all $t$ below some threshold, so the stability condition in \Eq{Re_cond} holds as $t\rightarrow-\infty$.
This implies that ${\rm Re}(\nuQ_\alpha)<0$ for all the roots of $Q$, so the Regge-like sums in Eq.~\eqref{eq:Ahard} become 
\eq{
\lim_{t\rightarrow\infty}\sum_{\alpha\neq 0} \AV(\nuQ_\alpha,t) &\sim \lim_{t\rightarrow\infty}\sum_{\alpha\neq 0} t^{\nuQ_\alpha} &&\hspace{-4mm}=0\\
\lim_{s\rightarrow\infty} \sum_{\beta\neq 0} \AV(s,\nuQ_\beta)&\sim \lim_{s\rightarrow\infty}\sum_{\beta\neq 0} s^{\nuQ_{\beta}}&&\hspace{-4mm}=0.
}{}
Thus, in the physical regime of high-energy fixed-angle scattering, our bespoke amplitude limits to a constant,
\be  
\AH(s,t) = \sum_{\alpha,\beta \neq 0} \AV(\nuQ_\alpha,\nuQ_\beta)\qquad \textrm{for} \qquad t<0.\label{eq:Ahard2}
\ee
In principle, it may be possible to choose the spectral curve such that the right-hand side is zero.  In this case, the resulting amplitude would exponentially vanish in the physical limit of high-energy fixed-angle scattering.

\subsubsection{Low-Energy Limit}

Last but not least, let us compute the amplitude at low energies, corresponding to  the leading behavior as $s,t\,{\rightarrow}\,0$.   In this limit, the roots of the spectral curve asymptote to the roots of $P$ and are thus $s$-independent.  Taking the low-energy limit, we obtain the effective field theory amplitude,
\eq{
\AEFT(s,t)  &= \! \lim_{s,t\rightarrow0}\!  A(s,t)
= \sum_{\alpha,\beta} \AV(\nuP_\alpha,\nuP_\beta),
}{eq:AIR}
dropping terms that are higher order in the derivative expansion.
Note that \Eq{eq:AIR} is only sensible if none of the roots $\nuP_\alpha$ are nonnegative integers.  Otherwise, the expression involves a gamma function evaluated on a singularity, yielding a formal divergence.      

We must therefore treat separately the physically interesting case in which our spectral curve produces a massless state.  In particular, let us assume that there is a single zero in the spectrum, $\mu(n_*) = 0$ for some level $n_*$, in which case there is a vanishing root, $\nuP_{*} = 0$.  Let us further assume that the spectral curve does not have double poles, so this zero root is unique. 
Expanding about zero, we have $\nu_*(s) = 0 + s \nu_*'(0) + \frac{1}{2}s^2 \nu_*''(0)+\cdots$, and we use the identity for derivatives of inverse functions to write $\nu_*'(0)=1/\mu'(n_*)$ and $\nu_*''(0) = -\mu''(n_*)/(\mu'(n_*))^3$.
 Assuming that none of the remaining roots $\nuP_{\alpha\neq  *}$ are nonnegative integers, we obtain the low-energy expansion for the amplitude when there is a massless mode,
\be 
\begin{aligned}
\AEFT = &-\h\mu'(n_*) \left(\frac{1}{s}+ \frac{1}{t}\right) \\& -\frac{\h \mu''(n_*)}{\mu'(n_*)} - 2\sum_{\alpha\neq *} H(-1-\nuP_\alpha) + \sum_{\alpha,\beta\neq *}\AV(\nuP_\alpha,\nuP_\beta)  \\& +\cdots  ,
\end{aligned}
\label{eq:AIR2}
\ee
where $H(n) = \gamma+\psi(n+1)$ is the analytically continued harmonic number, and the ellipses denote higher-derivative corrections.  The first line of \Eq{eq:AIR2} describes the exchange of a massless scalar.  On the other hand, the second line shows that the low-energy expansion generically includes a quartic contact term, which thus differs from the Veneziano amplitude.

\subsection{Dual Resonance}

As long as the roots of the spectral curve satisfy the negative real condition in \Eq{Re_cond}, then the Regge limit of the amplitude will be convergent.  In this case we can compute the bespoke dual resonant amplitude by computing the Galois sum on the dual resonant representation of the Veneziano amplitude directly, as shown in \Eq{eq:Galoisfull_again}.  

As before, we can use the fundamental theorem of algebra to write the Galois sum of the $s$-dependent part of  \Eq{eq:Galoisfull_again} as a $\dlog$ form, 
\eq{
\sum_{\alpha} \frac{1}{n-\nu_\alpha(s)} &= \partial_n \log f(s,n)= \frac{\partial_n f(s,n)}{f(s,n)}.
}{s-dep}
Crucially, by performing the Galois sum in this way, we do not have to explicitly derive the roots of the spectral curve.  This is not merely a convenience.  Indeed, the Abel-Ruffini theorem famously implies that it is {\it literally impossible} to construct a closed-form expression for the roots of quintic or higher polynomials in terms of radicals.  Plugging the formula for the spectral curve in \Eq{eq:singlebranch} into \Eq{s-dep}, we obtain a more explicit expression,
\eq{
\sum_{\alpha} \frac{1}{n-\nu_\alpha(s)}  &= \frac{\mu'(n)}{\mu(n)-s}  + \frac{Q'(n)}{Q(n)},
}{s-dep_simp}
which by construction exhibits a simple pole at $s=\mu(n)$, together with a contact term.  

The normalization of the simple pole fixes the normalization of the residue of the bespoke dual resonant amplitude in \Eq{eq:Galoisfull_again}, so 
\eq{
R(n,t) &= \lim_{s\rightarrow \mu(n)}(\mu(n)-s)A(s,t)
= 
\mu'(n) \sum_\beta R_V(n,\nu_\beta(t)) .
}{DR_res}
The Galois sum of the $t$-dependent residue in \Eq{DR_res} is much more complicated to evaluate.   In what follows, we describe two independent methods for performing this calculation.  In the first approach, we evaluate the sum using Newton's identities, while the second utilizes Cauchy's theorem.
We will find that the residues $R(n,t)$ are indeed polynomials in $t$ and thus consistent with locality.

\subsubsection{Residue from Newton's Identities}\label{sec:Newton}

Plugging \Eq{eq:Stirling} into the residue of the dual resonant amplitude defined in \Eq{DR_res}, we obtain the explicit formula,
\eq{
R(n,t)  & = \frac{\mu'(n)}{n!}\sum_{k=0}^n \begin{bmatrix} n+1 \\ k+1 \end{bmatrix} d_k(t),
}{DR_res_GN}
where we have defined the power sum of roots,
\eq{
d_k(t) = \sum_{\alpha} (\nu_\alpha(t))^k.
}{}
At this stage we invoke a powerful result from Galois theory.  In particular, the fundamental theorem of symmetric polynomials says that any polynomial that is symmetric in its variables can itself be written as a polynomial in a set of {\it elementary} symmetric polynomials.  Applied to the present context, we interpret each power sum $d_k(t)$ as a polynomial in the variables defined by the roots $\nu_\alpha(t)$.  Consequently, any power sum can be rewritten in terms of the elementary symmetric polynomials $e_k(t)$, which are defined by
\eq{
e_0(t)  \; &= \;\;\;\; 1 \\
e_1(t)   \; &=\;\;\;\,\, \sum_{\alpha_1}\;\;\; \nu_{\alpha_1}(t) \\
e_2(t)  \;  &= \;\;\, \sum_{\alpha_1 <\alpha_2} \,\, \nu_{\alpha_1}(t)\nu_{\alpha_2}(t) \\
e_3(t)  \;  &= \sum_{\alpha_1 <\alpha_2<\alpha_3} \!\! \nu_{\alpha_1}(t)\nu_{\alpha_2}(t)\nu_{\alpha_3}(t) \\
&\;\; \vdots \\
e_n(t)  \;  &=\;\;\;\; \nu_{\alpha_1}(t)\nu_{\alpha_2}(t) \cdots\nu_{\alpha_n}(t) ,
}{}
where $e_k(t)=0$ for $k>\h$.   
Since the power sum $d_k(t)$ is a symmetric polynomial, we can write it in terms of elementary symmetric polynomials $e_k(t)$ using Newton's identities, 
\eq{
d_k &= \sum_{i=k-\h}^{k-1} (-1)^{k-1+i} e_{k-i} d_i &&\quad \textrm{for}\quad k>\h\\
d_k &=(-1)^{k-1} k e_k +  \sum_{i=1}^{k-1} (-1)^{k-1+i} e_{k-i} d_i && \quad \textrm{for} \quad k\leq \h .
}{GNid}
Conveniently, even though Newton's identities are defined recursively, they have a closed-form solution known as the Girard-Waring formula, given in determinant form in Ref.~\cite{Gould},
\eq{ d_k(t) = (-1)^k
\left|\begin{matrix}
-e_1(t) & +e_0(t) & 0 & {\cdots} & 0\\
+2e_2(t)&-e_1(t) & +e_0(t) & {\cdots} & 0\\
-3e_3(t) & +e_2(t) & -e_1(t) & {\cdots} & {\vdots}\\
{\vdots} & {\vdots} & {\vdots} & {\ddots} & +e_0(t)\\
(-1)^k k e_k(t) & (-1)^{k-1} e_{k-1}(t) & (-1)^{k-2} e_{k-2}(t)& {\cdots}  & -e_1(t)
\end{matrix}\right|.
}{GNid2}
The coefficients of any monic polynomial are directly related to the elementary symmetric polynomials built from the roots of that polynomial, specifically,
 \eq{
f(t,\tau)&= \sum_{k=0}^\h (-1)^{\h-k} e_{\h-k}(t) \tau^k  .
}{fexp}
This identity implies that the elementary symmetric polynomials can be extracted directly from the spectral curve via 
\eq{
e_k(t) &= \frac{(-1)^{k}}{(\h-k)!} \partial_\tau^{\h-k} f(t,\tau) \big|_{\tau=0}.
}{ek_def}
Consequently, the residue of the bespoke dual resonant amplitude in  \Eq{DR_res_GN} can be explicitly evaluated by writing the power sums $d_k(t)$ in terms of the coefficients of the spectral curve via \Eq{GNid2} and \Eq{ek_def}.     By definition, $f(t,\tau)$ is a monic polynomial in $\tau$ with coefficients that are themselves linear polynomials in $t$. As such, \Eq{ek_def} then implies that the functions $e_k(t)$ are also polynomials in $t$, and hence so too is $d_k(t)$, on account of \Eq{GNid2}.

Plugging into \Eq{ek_def} our explicit formulas for the spectral curve in Eqs.~\eqref{eq:singlebranch} and \eqref{eq:PQ}, we obtain a simple expression for the elementary symmetric polynomials,
\eq{
e_k(t) = (-1)^k(p_k - t q_k),
}{ek_exp} 
where $q_0 = 0$. Since $e_k(t)$ is of degree one, \Eq{GNid2} implies that  $d_k(t)$ is a polynomial of degree $k$.  Hence, we learn that \Eq{DR_res_GN}, the residue $R(n,t)$, is a {\it polynomial} in $t$ of degree $n$, corresponding to a maximal spin of $n$ on the pole at $s=\mu(n)$. 
This is an important check, since the presence of residues that are not exponentially suppressed or zero at arbitrarily high spin is a sign of nonlocality~\cite{Huang:2022mdb,Caron-Huot:2020cmc}, which we wish to avoid.

\subsubsection{Residue from Cauchy's Theorem}

There is an alternative method for computing the residue of the dual resonant amplitude defined in \Eq{DR_res} using Cauchy's theorem.
In particular, \Eq{DR_res} can be recast into the form of a $\dlog$ integral,
\eq{
R(n,t) &= \frac{1}{2\pi i}\mu'(n) \oint \dlog(f(t,\tau)) \RV(n,\tau)\\
&=-\mu'(n) \Res{\tau = \infty}\big[ R_V(n,\tau)  \partial_\tau \log f(t,\tau)   \big],
}{}
where, in the first line, the contour of integration is the sum of loops encircling the zeroes of the spectral function.   In the second line, we have applied Cauchy's theorem to blow up that contour to the boundary at infinity.

By expanding the residue at infinity as a series in $t$, we can write the residue as an explicit polynomial,
\eq{
R(n,t) = \sum_{k=0}^n b_{k}(n) t^k,
}{res_Cauchy1}
where the sum truncates at $k\,{=}\,n$ since $R(n,t)$ has degree at most $n$, as follows from the definition of the spectral curve in  Eqs.~\eqref{eq:singlebranch} and \eqref{eq:PQ}.
The coefficients of the residue polynomials are given by the closed-form expression
\eq{
b_k(n) &= -\mu'(n) \Res{\tau = \infty}\big[ R_V(n,\tau)\left.\partial_\tau \partial_t^k\log f(t,\tau)\right |_{t=0} \;\big] \\
&= \frac{(-1)^{n-k} \mu'(n)}{k!(n-k)!}\lim_{\tau\rightarrow\infty} \left[\tau^{n-k+1}\partial_\tau^{n-k}\left(R_V(n,\tau) \left.\partial_\tau \partial_t^k \log f(t,\tau)\right|_{t=0}\right) \right],
}{res_Cauchy2}
which yields the same results computed using Newton's identities in the previous section.

\subsubsection{Summary of Results}

For later convenience, let us briefly summarize the results of the above calculation.  We have evaluated the Galois sums in the dual resonant form of the amplitude in \Eq{eq:Galoisfull_again} to obtain 
\eq{
A(s,t)&=  \sum_{n=0}^\infty  \left(\frac{1}{\mu(n) - s} + \frac{Q'(n)}{\mu'(n)Q(n)}\right) R(n,t).
}{Afinal1}
Here the $s$-dependent factor was computed in \Eq{s-dep_simp}, while the $t$-dependent residue was computed via Newton's identities in Eqs.~\eqref{DR_res_GN}, \eqref{GNid2}, and \eqref{ek_exp}, and via Cauchy's theorem in Eqs.~\eqref{res_Cauchy1} and \eqref{res_Cauchy2}.

Since the second term in \Eq{Afinal1} is independent of $s$, it corresponds to the boundary term in the unsubtracted dispersion relation in \Eq{eq:SDR}, so
\eq{
A_\infty(t) = \sum_{n=0}^\infty \frac{R(n,t) Q'(n)}{\mu'(n)Q(n)} .
}{Afinal2}
Alternatively, we can also reabsorb this boundary term into the numerator by defining an $s$-dependent residue.  In particular, we can write the bespoke dual resonant amplitude as
\eq{
A(s,t)&=  \sum_{n=0}^\infty  \frac{R(n,s,t)}{\mu(n) - s}  \qquad\textrm{where} \qquad 
R(n,s,t) = R(n,t) \left[1 + \frac{(\mu(n) - s)Q'(n)}{\mu'(n)Q(n)}\right].
}{Afinal3}
Here $R(n,s,t)$ and $R(n,t)$ are equal on the pole at $s=\mu(n)$.  Thus we arrive at a central conclusion of this paper: a closed-form, dual resonant representation for a new infinite class of scattering amplitudes with a customizable spectrum.

Note that for the case of the Kaluza-Klein amplitude, the spectral curve in \Eq{spectral_KK} corresponds to $P(\nu) =(\nu+\delta)^2$ and $Q(\nu)=1$, for which the contact and boundary terms in Eqs.~\eqref{Afinal1}, \eqref{Afinal2}, and \eqref{Afinal3} all vanish.  In this case, these formulas accord with those derived earlier for the Kaluza-Klein amplitude.

\subsubsection{Constraint of Polynomial Residues}

Let us briefly comment on the constraint of polynomial residues, which is simply the requirement that the residue on each pole, $R(n,t)$, is a polynomial in $t$.  When this condition fails, the exchanged mode exhibits partial waves of arbitrarily high spin, indicating some intrinsic nonlocality in the theory.  

In our discussion thus far, we have focused on the case in which the spectral curve, defined in \Eq{eq:singlebranch}, is monic.  This requires that $|P| > |Q|$, in which case the residues $R(n,t)$ are polynomial.  What happens when $|P|\leq |Q|$?

It is easiest to see why this other choice is pathological with a concrete example.   To this end, let us consider an {\it accumulation point spectrum}, defined by $f(\mu,\nu)= 1-\mu \nu$, corresponding to $P(\nu)=1$ and $Q(\nu)=\nu$.  In this case, the spectrum is $\mu(\nu) = 1/\nu$, which accumulates to zero.  For this choice of spectrum, the dual resonant amplitude in \Eq{eq:M} effectively maps $s\rightarrow 1/s$ and $t\rightarrow 1/t$ in the Veneziano amplitude.  This leads to residues that are not polynomials in $t$, but rather in $1/t$, violating locality.\footnote{A similar issue arises for the Coon amplitude, where a log-dependent prefactor cures the inverse powers of $t$ in the residues, though at the expense of introducing potential issues with unitarity~\cite{Jepsen}. While it would be interesting to see if similar prefactors can allow for consistent accumulation-point amplitudes using our Galois sum construction, we will not pursue this idea here.}

Our constructions using Newton's identities or Cauchy's theorem do not apply to the spectrum described above, simply because the spectral curve is not monic: the highest power of $\nu$ in $f(\mu,\nu)$ has coefficient $-\mu$. Hence, while the fundamental theorem of symmetric polynomials guarantees that the $R(n,t)$ are polynomial in the elementary symmetric polynomials $e_i$, the violation of the monic condition implies that the $e_i$ are no longer simply a remapping of the coefficients of the spectral curve up to signs as given in Eq.~\eqref{fexp}.
The residue $R(n,t)$ therefore need not be polynomial for spectral curves that are not monic.

\section{Examples}\label{sec:examples}

We now turn to some concrete examples in order to illustrate how our construction works in explicit detail.   In particular, for these cases we will compute the residues and asymptotics for the corresponding bespoke amplitudes and determine the parameter regions that are consistent with partial wave unitarity.  The latter requirement mandates positivity of the coefficient $a_{n,\ell}\geq 0$ of the spin-$\ell$ partial wave of the resonance at level $n$.  For the sake of brevity, we will not recapitulate the familiar story of unitarity bounds, but refer the interested reader to a review of those details in App.~\ref{app:unitarity}.

\subsection{Simplest Nonlinear Model}

The degree $h$ of the spectral curve in \Eq{eq:singlebranch} bounds the maximal complexity of the spectrum.  For $\h=1$, the spectral curve is linear, and our construction simply corresponds to an affine transformation acting on the kinematic arguments of the Veneziano amplitude.  Hence, the first nontrivial case occurs when spectral curve becomes nonlinear at $\h=2$, which we now consider.  
In this case, the spectral curve is defined by the following polynomials:
\eq{
P(\nu) = \nu^2 + p_1 \nu+ p_2 \qquad \textrm{and} \qquad
Q(\nu) = \nu + q_2.
}{}
This class of models has asymptotically linear Regge behavior, since $\lim
\limits_{\nu\rightarrow\infty} \mu(\nu)/\nu=1$, but exhibits a nonlinear spectrum at finite $\nu$. 

The zero locus of the spectral function, $f(\mu,\nu)=0$, defines the roots
\be
\nu_{\pm}(\mu) = \frac{1}{2}\left(\mu - p_{1}{\pm}\sqrt{p_{1}^{2}-4p_{2}-2(p_{1}-2q_{2})\mu+\mu^{2}}\right),
\ee
so the corresponding bespoke dual resonant amplitude is given by \Eq{eq:M}.  In order to ensure convergence of the asymptotic behavior, 
we impose the negative real condition in \Eq{Re_cond}, so $\textrm{Re}\left(\nu_{\pm}(t)\right) <0$.  Demanding this constraint for all $t<0$ implies that
\be
p_1,p_2,q_2>0. \label{eq:stronger}
\ee
On the other hand, if we merely demand that there exists {\it some} real value of $t$ for which the asymptotics are convergent, then we obtain 
\be
q_2 > 0 \qquad \textrm{or} \qquad p_{2}>p_{1}q_{2},\label{eq:weaker}
\ee
which is a weaker sufficient condition.  For any choice of parameters conforming with these conditions, the scattering amplitude in \Eq{eq:M} will have a dual resonant representation. 

\begin{figure}[t]
\begin{center}
\includegraphics[width=\textwidth]{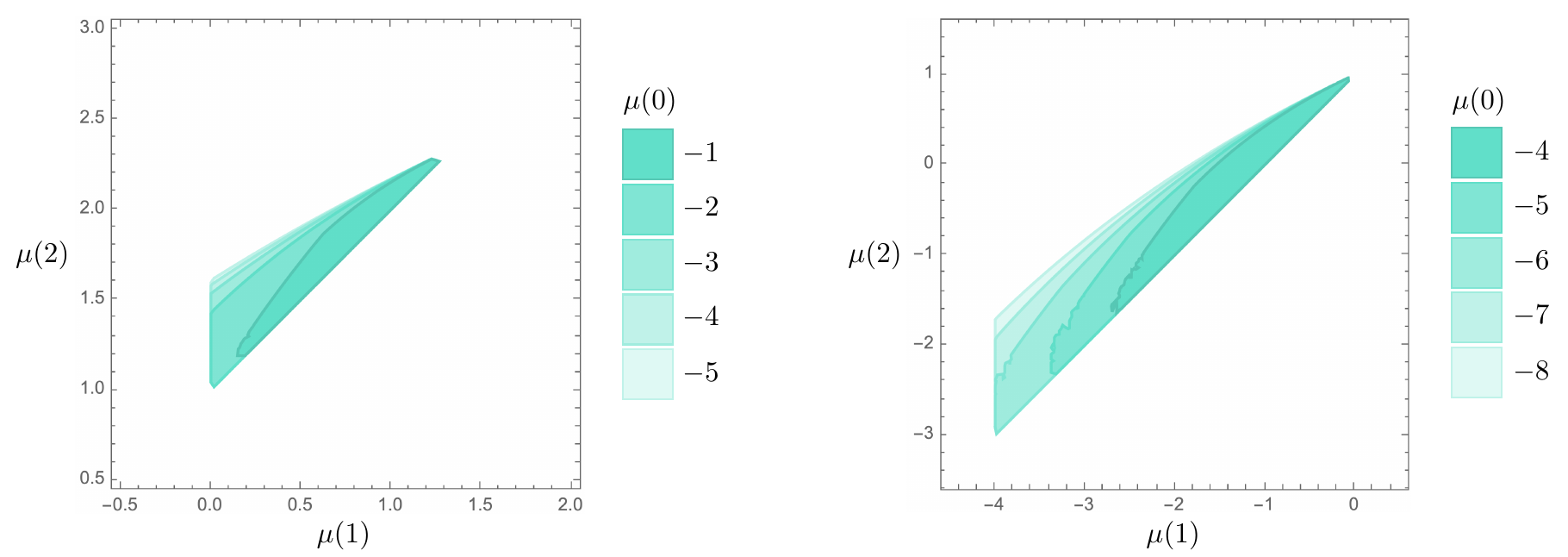}
\end{center}
\vspace{-0.5cm}
\caption{Parameter regions consistent with partial wave unitarity and the weaker dual resonance condition in \Eq{eq:weaker} for the simplest nonlinear models.  Here, the left and right panels correspond to $m_{\rm ext}^2 = 0$ and $-1$, respectively, working in spacetime dimension $D=4$. We have displayed constraints from all partial waves through $n\leq  6$, beyond which the allowed region does not appreciably change. The value of the lightest particle, $\mu(0)$, is apparently not constrained by unitarity for arbitrarily negative values.
}
\label{fig:h2}
\end{figure} 

Alternatively, we can employ a more physical parameterization in which the variables $p_1$, $p_2$, and $q_2$ are remapped onto the values of the first three masses in the spectrum, $\mu(0)$, $\mu(1)$, and $\mu(2)$.  We can then impose the constraints of partial wave unitarity on the physical mass spectrum.  In Fig.~\ref{fig:h2}, we have computed the range of masses consistent with partial wave unitarity up through $n\leq 6$, together with the weaker dual resonance condition in \Eq{eq:weaker}.  Here we have considered the cases in which the mass squared for the external states is $m_{\rm ext}^2 = 0$ or $-1$.

Among the simplest nonlinear models, there is a particularly nice choice of parameters, $p_1 = p_2 = \delta$ and $q_2 = 1$, which yields a scattering amplitude with several exceptional properties. For these parameters, we obtain a spectrum of the form
\be
\mu(\nu) = \frac{\nu^2}{\nu+1} + \delta, 
\ee 
which enjoys the  curious duality invariance,
\eq{ 
\nu+1 \leftrightarrow \frac{1}{\nu+1}.
}{}
The conditions in Eqs.~\eqref{eq:stronger} and \eqref{eq:weaker} imply that dual resonance holds for all $t<\textrm{min}(0,\delta)$.  Plugging these special parameters into \Eq{eq:Galasymp}, we obtain an incredibly simple, $t$-independent expression for the boundary term at infinity, 
\be
A_\infty(t) = \sum_{\beta} A_V(-1,\nu_\beta(t)) = -\sum_{\beta} \frac{1}{\nu_\beta(t)} = 1,
\ee
which also agrees with numerical evaluation of the Regge limit.
As one can check, this result agrees with Eq.~\eqref{Afinal2}, with the factor of $\mu'(n)$ in the denominator cancelling that in $R(n,t)$ in Eq.~\eqref{DR_res_GN}.
Including this contact term, we obtain the corresponding amplitude,
\be 
A(s,t) = 1 + \sum_{n=0}^\infty \frac{R(n,t)}{\mu(n)-s},
\ee
which can be written in the dual resonant form in \Eq{Afinal3} simply by reabsorbing the contact term into the $s$-channel sum.

For this special model, we have computed the bounds from particle wave unitarity up to level $n\leq 40$, assuming $m_{\rm ext}^2 = 0$ in $D=4$ dimensions.  These constraints imply that $-1/2\leq \delta \lesssim -0.3541$.
Remarkably, for this family of amplitudes, the state at $n=0$ has vanishing partial wave coefficient, $a_{0,0}=0$, so there is no associated pole, and the $n=0$ state can be excised from the spectrum.
In contrast, at level $n\geq 1$ the associated partial waves $a_{n,\ell}$ for $0\leq \ell \leq n$ are all nonzero.  Finally, we note that the particular choice of $\delta=-1/2$ is interesting, since it sets $\mu(1)=0$, so the lightest exchanged states are a massless scalar and vector, where the former can in principle be identified with the external state.

\subsection{Post-Regge Expansion}

Next, let us consider a family of amplitudes defined by a spectral curve of arbitrary degree $\h$.  In order to reduce the parameter space to a manageable size, we will assume a particularly simple form for the spectral curve, defined by
\eq{
P(\nu) = \sum_{i=0}^\h \kappa_i (\nu-\nu_*)^{\h-i} \qquad \textrm{and} \qquad Q(\nu) = (\nu -\nu_*)^{\h-1} ,
}{}
where $\kappa_0=1$ to keep the spectral curve monic.
From the discussion in Sec.~\ref{sec:control}, we saw that dual resonance for some sufficiently negative value of $t$ requires that the roots of $Q$ all have negative real components.  This is ensured provided we assume that $\nu_*<0$. Note that $\kappa_0=1$ so that the spectral curve is monic.

By construction, the spectrum of this class of theories is controlled by the function
\be
\mu(\nu) = \frac{P(\nu)}{Q(\nu)}=\sum_{i=0}^\h \kappa_i (\nu-\nu_*)^{1-i}  = (\nu - \nu_*) + \kappa_1 + \frac{\kappa_2}{\nu-\nu_*} + \frac{\kappa_3}{(\nu-\nu_*)^2} + \cdots,
\ee
describing a series expansion in inverse powers of $\nu-\nu_*$.  The leading term is simply the asymptotic linear Regge behavior familiar from string theory.  However, our construction now allows for customizable post-Regge {\it corrections} to this leading behavior.  For this reason, we will refer to this spectrum as a post-Regge expansion.

Let us now turn to the question of partial wave unitarity. 
A priori, these amplitudes have $\h+1$ free parameters given by $\nu_*$ and $\kappa_{i}$.  
The enormous size of this parameter space, even for modest values of $\h$, will require ad hoc choices of parameters simply in order to visually depict any bounds.   For this reason we will cut down the parameter space to $\h-1$ variables by imposing the constraint that $a_{0,0} = a_{1,1} = 0$.  This choice is not entirely random.  Rather, it has the advantage of completely excising the $n=0$ resonance from the spectrum $\mu(n)$, allowing us to contemplate large negative $\mu(0)$ values with impunity.  Furthermore, it removes the $\ell=1$ partial wave at $n=1$, so we are safe from spinning tachyons at that level~\cite{Thaler,Dubovsky:2006vk}.  We will assume this choice of parameters for the remainder of this section.

\begin{figure}[t]
\begin{center}
\includegraphics[width=\textwidth]{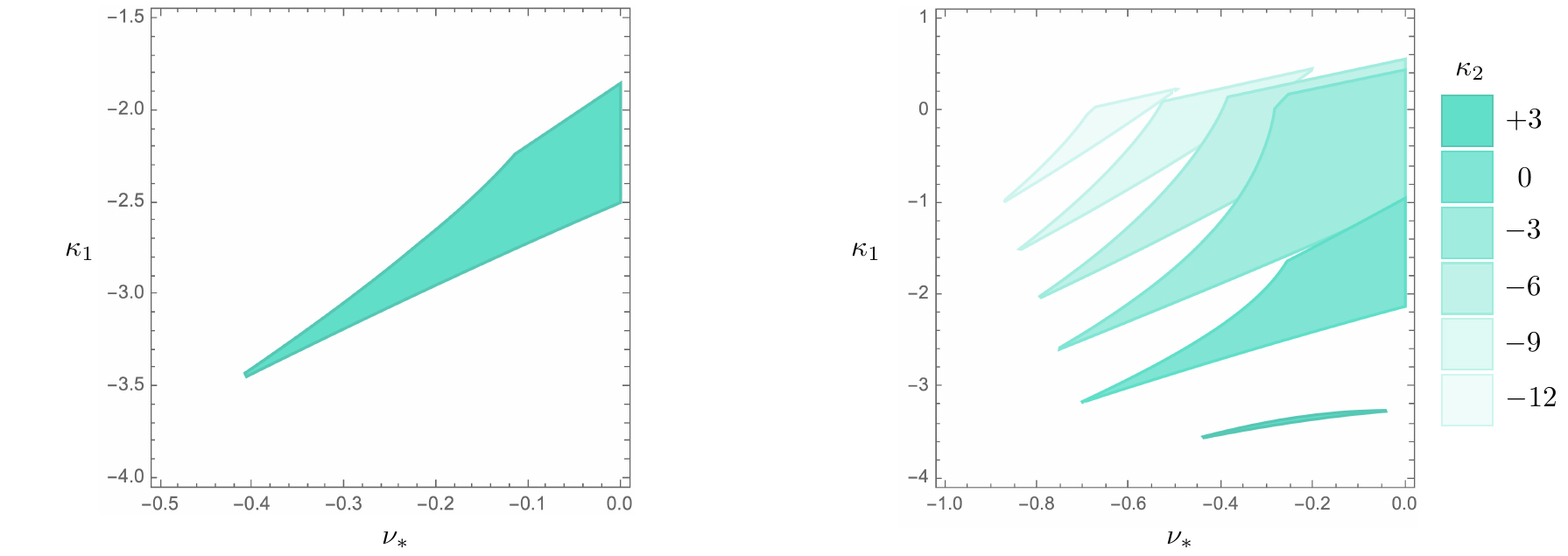}
\end{center}
\vspace{-0.5cm}
\caption{Parameter regions of the amplitudes in the post-Regge expansion consistent with partial wave unitarity up to $n\leq 8$ for the $\h=3$ (left) and $\h = 4$ (right) models, where $\kappa_{\h-1}$ and $\kappa_{\h}$ are fixed by requiring $a_{0,0} = a_{1,1} = 0$.   Furthermore, we have set $m_{\rm ext}^2 =0$ and $D=4$.}
\label{fig:PR}
\end{figure} 

There are many ways to visualize the bounds on this restricted parameter space.  In our first approach, we consider the case in which we set the external masses to $m_{\rm ext} = 0$ and the spacetime dimension to $D=4$, while eliminating $\kappa_{\h-1}$ and $\kappa_{\h}$ via the conditions $a_{0,0} = a_{1,1} = 0$.
We then compute the constraints from partial wave unitarity up to $n\leq 8$. 
For $\h=2$, this leaves a one-parameter space of theories, constrained by $-1.2293\lesssim \nu_* < 0$.   Meanwhile, for $\h=3$ and $\h=4$, we plot the regions consistent with partial wave unitarity in Fig.~\ref{fig:PR}.

\begin{figure}[t]
\begin{center}
\includegraphics[width=7cm]{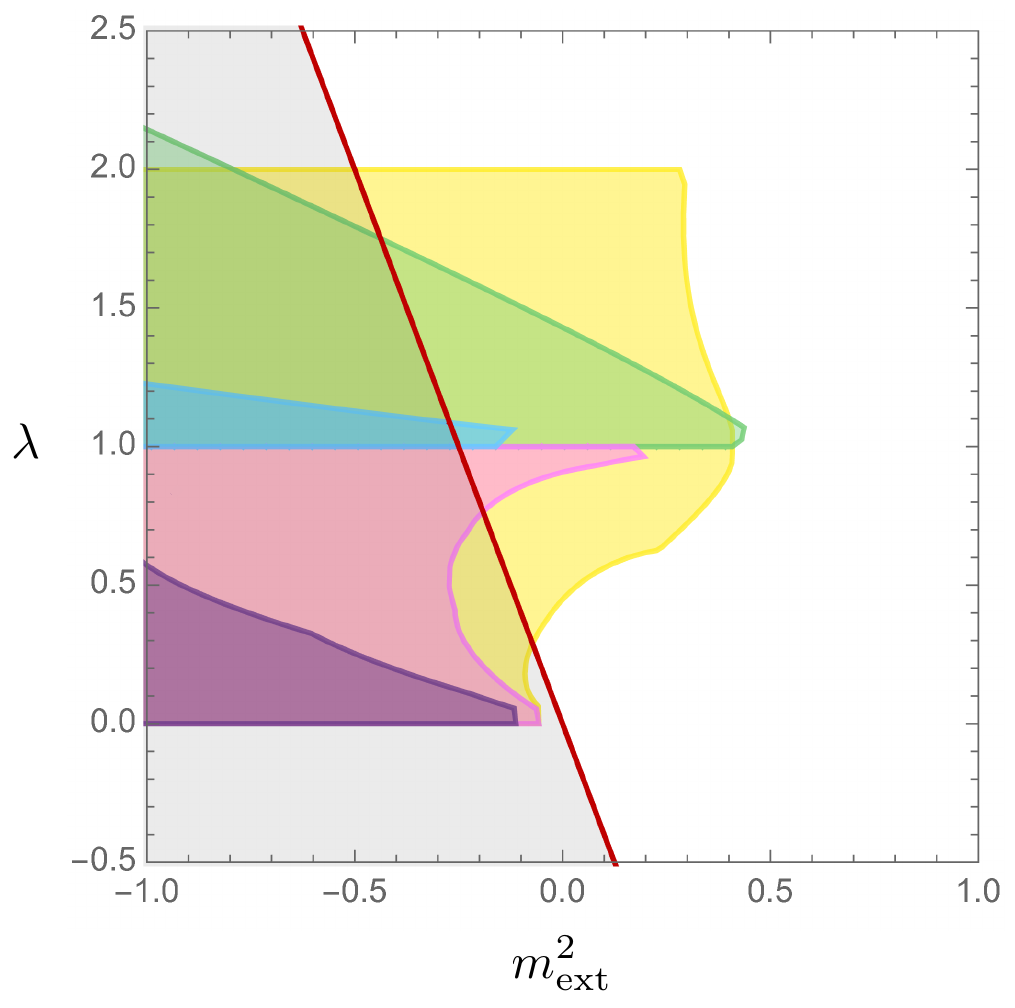}
\end{center}
\vspace{-0.5cm}
\caption{Parameter regions of the amplitudes in the post-Regge expansion consistent with partial wave unitarity up to $n\leq 5$ for the $\h=2$ (yellow), $\h=3$ (green), $\h=4$ (pink), $\h=5$ (blue), and $\h=6$ (purple) models.  Here we fix $\nu_*$ and $\mu(0)$ via $a_{0,0} = a_{1,1} = 0$ and set $D=4$ and $\mu(\nu) = \lambda \nu + \mu(0)$ for $\nu = 0,\ldots,\h-1$.
The shaded gray region to the left of the red line is forbidden by requiring that $\mu(2) \geq 0$ in order to avoid spinning tachyons.}
\label{fig:PRm}
\end{figure}

In a second approach, we rewrite the $\h+1$ free parameters of the spectral curve in terms of the first $\h$ masses in the spectrum, along with $\nu_*$.  In principle, these physical masses can be freely chosen.  However, again to reduce the parameter space to a reasonable size, we constrain the first $\h$ masses to a line of arbitrary slope and intercept, so
\eq{
\mu(\nu) = \lambda \nu + \mu(0) \qquad \textrm{for} \qquad \nu = 0,\ldots,\h-1.
}{}
We now eliminate $\nu_*$ and $\mu(0)$ via the conditions $a_{0,0} = a_{1,1} = 0$. 
In the end, we obtain a family of amplitudes controlled by just two parameters, $\lambda$ and $m_{\rm ext}^2$.  These variables are particulary nice because they transparently characterize how much the low-lying states can deviate from a linear spectrum with unit slope.
In particular, Fig.~\ref{fig:PRm}  depicts the regions of parameter space consistent with partial wave unitarity up to $n\leq 5$ for $\h=2,3,4,5,6$.  Notice that $\lambda$ is restricted to a relatively narrow range of values, demonstrating that partial wave unitarity disfavors large deviations from the linear Regge spectrum of string theory.
This apparent stiffness under the simultaneous constraints of dual resonance and partial wave unitarity comports with the results of asymptotic uniqueness~\cite{Zohar}.

\section{Higher-Point Scattering}

Up to this point, our analysis has focused solely on the case of four-point scattering.  Perhaps surprisingly, our construction has a mechanical generalization to higher-point scattering, which we now discuss.  In particular, the resulting higher-point amplitudes have a customizable spectrum, exhibit an integral representation reminiscent of a worldsheet, and pass some basic consistency checks on factorization.  Nonetheless, we emphasize here that the full consistency of these higher-point bespoke scattering amplitudes is far from established.

\subsection{Remapping the Spectrum}

As a warmup, let us revisit the higher-point string amplitudes, which we denote by $\AV(\left\{ s_I \right\})$.  These objects are most easily formulated using the Koba-Nielsen formula~\cite{KobaNielsen}.  Here $\{ s_I\}$ denotes the minimal basis of {\it planar} kinematic invariants, which are labeled by an index $I$ running over choices of contiguous sets of the external momentum labels. For example, for low-point amplitudes this basis is
\eq{
\textrm{four-point:} &  \quad\{ s_{12}, s_{23} \} \\
\textrm{five-point:} & \quad\{ s_{12}, s_{23},s_{34},s_{45},s_{51} \} \\
\textrm{six-point:} & \quad\{  s_{12}, s_{23},s_{34},s_{45},s_{56},s_{61}, s_{123}, s_{234}, s_{345} \} ,
}{}
and so on, where $s_{i_1 \cdots i_k} = -(p_{i_1} + \cdots + p_{i_k})^2$.  In general, for the scattering of $N$ particles, the minimal planar basis will have $N(N-3)/2$ invariants.  See Ref.~\cite{TASI} for more details.
The advantage of utilizing a basis of planar invariants follows from the fact that planar amplitudes only exhibit factorization channels among adjacent sets of legs.  As a result, factorization in a channel labeled by $I$ simply corresponds to setting $s_I$ equal to the mass squared of the corresponding intermediate state.

To generalize to higher-point scattering, we still define a spectral curve $f(\mu,\nu)$ of the form in Eqs.~\eqref{eq:singlebranch} and \eqref{eq:PQ}, whose zeros dictate the spectrum $\mu(\nu)$ of the theory, as well as a family of inverse roots $\nu_\alpha(\mu)$.  It is then natural to define a new higher-point bespoke amplitude,
\eq{
A(\{ s_I\}) = \left( \prod_I \sum_{\alpha_{I}} \right) \AV(\{ \nu_{\alpha_I}(s_I)\}),
}{eq:Mn}
which simply folds each planar invariant $s_I$ with a function $\nu_{\alpha_I}(s_I)$, and then sums over all branches of roots labeled by $\alpha_I$.  To be very explicit, at four-point \Eq{eq:Mn} yields our original formula from \Eq{eq:M},
\eq{
A(s_{12}, s_{23}) &= \sum_{\alpha_{12},\alpha_{23}} \AV(\nu_{\alpha_{12}}(s_{12}), \nu_{\alpha_{23}} ( s_{23})),
}{4ptsum}
while at five-point we obtain the analogous expression,
\eq{
\! A(s_{12}, s_{23},s_{34},s_{45},s_{51})  = \! \sum_{\substack{\alpha_{12},\alpha_{23},\alpha_{34},\\ \alpha_{45},\alpha_{51}}} \! \AV(\nu_{\alpha_{12}}(s_{12}), \nu_{\alpha_{23}} ( s_{23}),\nu_{\alpha_{34}} ( s_{34}),\nu_{\alpha_{45}} ( s_{45}),\nu_{\alpha_{51}} ( s_{51})),
}{5ptsum}
and so on.  As before, we can recast the sum over roots into the form of a $\dlog$ contour integral over the higher-point string amplitude,
\eq{
A(\{ s_I\}) = \left( \prod_I \oint \frac{\dlog f(s_I,\sigma_I)}{2\pi i} \right) \AV(\{ \nu_{\alpha_I}(\sigma_I)\}),
}{}
where each contour of integration is simply given by the sum of loops wrapping the poles defined by $f(s_I,\sigma_I)=0$.

Since the higher-point bespoke amplitude in \Eq{eq:Mn} is simply a Galois sum of the higher-point Veneziano amplitude, the former will automatically inherit many properties of the latter.  One such property is invariance under the various relabeling symmetries of the original string amplitude.
For example, any cyclic permutation on the labels of the external particles will correspond to some permutation of the $s_I$ among themselves, which is in turn equivalent to the action of the {\it inverse} permutation on the $\alpha_I$, on account of the underlying cyclic symmetry of $A_V$ itself.
Since all possible choices of the $\alpha_I$ appear democratically in the Galois sum, this relabeling leaves the bespoke amplitude invariant as well. 
The same statements apply to the flip symmetry of the string amplitude. 

Another inherited property is dual resonance.  In particular, any dual resonant representation of the higher-point Veneziano amplitude~\cite{ChanTsou} can be inserted into the Galois sum to produce a corresponding dual resonant representation of the higher-point bespoke amplitude.  Then one need only check that this sum is convergent, which we suspect is ensured for amplitudes with sufficiently tame ultraviolet behavior.  We leave a full analysis of this important question to future work.

\subsection{Worldsheet Representations}\label{sec:worldsheet}

Given that the higher-point bespoke amplitude in \Eq{eq:Mn} is a sum over all roots of the spectral curve, it is natural to anticipate that it is free of branch cuts.  Indeed, we can verify that this is the case by constructing an integral representation for these objects.  As usual, our approach here will be to simply compute the Galois sum over the existing worldsheet representation for the higher-point string amplitudes.

To begin, recall the integral representation of the Veneziano amplitude,
\be
\AV (s_{12},s_{23}) = \int_0^1 \dr x\,x^{-s_{12}-1}(1-x)^{-s_{23}-1}, \label{eq:Aint}
\ee
which in string theory is derived from an integral over worldsheet moduli after gauge fixing.  Plugging this expression into \Eq{4ptsum}, we obtain the bespoke four-point amplitude,
\be
A(s_{12},s_{23}) = \int_0^1 \dr x \sum_{\alpha_{12}} x^{- \nu_{\alpha_{12}}(s_{12})-1}\sum_{\alpha_{23}} (1-x)^{-  \nu_{\alpha_{23}}(s_{23})-1}.\label{eq:Mint}
\ee
For convenience, let us define a special function,
\eq{
\rho(x,s) = \sum_{\alpha} x^{-\nu_{\alpha}(s)},
}{}
which is analytic in $s$ since it is manifestly a sum over roots.\footnote{One can verify that this is the case by expanding $x$ about any nonzero point, yielding a series whose coefficients are power sums $d_k(s)$ over the roots, which as before can be computed via Newton's identities.} Our next task will be to rewrite our amplitudes in terms of $\rho$.  In particular, the bespoke four-point amplitude is
\eq{
A(s_{12},s_{23}) = \int_0^1 \dr x \; \frac{\rho(x,s_{12}) \rho(1-x,s_{23})}{x(1-x)},
}{}
which is analytic in the kinematic invariants except at simple poles.  All branch cuts have cancelled in the above expression.
We emphasize that this cancellation was only possible because each planar kinematic invariant occurs in a distinct multiplicative factor in \Eq{eq:Aint}.

As one might anticipate, the same phenomenon occurs in the five-point string amplitude~\cite{Nakanishi:1971ve}, whose worldsheet representation can be written as
\eq{
&\AV(s_{12},s_{23},s_{34},s_{45},s_{51}) = \int_{0}^{1}\int_{0}^{1}\dr x\,\dr y\,\frac{x^{-s_{12}}\!\left(\frac{1-x}{1-xy}\right)^{-s_{23}} \! y^{-s_{45}}\!\left(\frac{1-y}{1-xy}\right)^{-s_{34}}\!(1\,{-}\,xy)^{-s_{51}}}{x(1-x)y(1-y)} .
}{}
We therefore have a worldsheet representation for the five-point bespoke amplitude,
\be 
\begin{aligned}
&A(s_{12},s_{23},s_{34},s_{45},s_{51}) \\&= \int_{0}^{1}\int_{0}^{1}\dr x\,\dr y \,\frac{\rho(x,s_{12})\rho\left(\frac{1-x}{1-xy},s_{23}\right)\rho(y,s_{45})\rho\left(\frac{1-y}{1-xy},s_{34}\right)\rho(1\,{-}\,xy,s_{51})}{x(1-x)y(1-y)}.
\end{aligned}
\ee
As before, the dependence on each planar kinematic invariant factorizes multiplicatively, in which case the amplitude can be written in a form that is manifestly free of branch cuts.  This factorization property occurs for any number of external particles.  To understand why, recall that any string amplitude is computed from a product of vertex operators that introduce kinematic dependence through factors of the form $(x_{i}-x_j)^{-s_{ij}}$, where $x_i$ and $x_j$ are moduli variables and the indices label external legs.  Crucially, we can always decompose any kinematic invariant into the planar basis, so
\eq{
s_{ij} = \sum_I c_{ijI} s_I.
}{}
As a result, we can recast any product of vertex operator contributions as
\eq{
\prod_{i<j} (x_{i}-x_j)^{-s_{ij}} =\prod_{i<j} (x_{i}-x_j)^{- \sum_I c_{ijI} s_I} = \prod_I \big[\prod_{i<j} (x_{i}-x_j)^{ c_{ijI} } \big]^{-s_I}.
}{}
The latter is manifestly a factorized product of terms, each of which depends on a different planar kinematic invariant.  Applying the sum over roots to a given factor on the right-hand side, we obtain
\eq{
\sum_{\alpha_I}\big[\prod_{i<j} (x_{i}-x_j)^{ c_{ijI} } \big]^{-\nu_{\alpha_I} (s_I)} = \rho\bigg(\prod_{i<j} (x_{i}-x_j)^{ c_{ijI} } , s_I\bigg),
}{}
which is expressed purely in terms of the analytic function $\rho$. All branch cuts thus cancel in our higher-point bespoke amplitudes.

We should pause to emphasize that while the above integral representations are strongly reminiscent of a worldsheet, we have not actually constructed a full theory in this regard.  In particular, it remains to be seen if there exist consistent vertex operators whose insertions generate the integral forms above.

\subsection{Factorization}

In our analysis of four-point scattering, we considered the constraint of partial wave unitarity but ignored factorization altogether.  This was not simply an omission.  Rather, the relevant three-point amplitudes are tautologically {\it defined} to be those which arise in the factorization channels of the four-point amplitude.  Thus, by construction factorization does not impose any meaningful constraint in going from four-point to three-point.  At higher-point, however, the condition of factorization is no longer automatic.

Proper factorization of higher-point string amplitudes---including exchanges of arbitrarily high mass and spin---is a mathematical property that has yet to be proven directly using amplitudes methods.    Instead, results in this area have centered primarily on factorization channels involving just the lightest states of the theory.  Here we will do the same and only consider factorization on the lowest-lying mode, which in our case is the single scalar residing at the $n=0$ level.  We leave the far more difficult question of factorization at general $n$ to future work.

For the higher-point Veneziano amplitude, factorization onto the massless scalar at level $n=0$ corresponds to $s_*=0$, where $s_*$ is the planar kinematic invariant corresponding to the factorization channel in question.  Consistency mandates that, on this factorization channel, the amplitude degenerates into a product of lower-point Veneziano amplitudes,
\eq{
\lim_{s_*\rightarrow 0 } s_* \,\AV(\{ s_I \}) = \sum_{L \perp R} \AV( \{ s_{I_L}\})\AV(\{ s_{I_R}\}).
}{eq:factorV}
Here, $L$ and $R$ are two mutually exclusive, contiguous subsets of the external labels whose union forms the full set of $N$ external particles.  The sum in \Eq{eq:factorV} runs over the partitions of all $N$ particles into the sets $L$ and $R$.

Meanwhile, the sets $ \{ s_{I_L}\}$ and $ \{ s_{I_R}\}$ denote the minimal bases of planar invariants for the lower-point amplitudes that appear in the factorization channel.  Crucially, for scalar exchange we know that the factorization channel cannot allow any dependence on kinematic invariants built from momenta that lie in different lower-point amplitudes. Hence the union of planar kinematic invariants that appear in the left and right amplitudes is a proper subset of the full set of planar kinematic invariants in the original amplitude, specifically, 
\eq{
 \{ s_{I_L}\} \cup  \{ s_{I_R }\} = \{ s_I \} \setminus \{s_*\},
}{}
where $s_*$ disappears on the factorization channel since it has been set to zero.  

Next, let us ask what happens when this same analysis is applied to the higher-point bespoke amplitude.  Here we consider the factorization channel onto the scalar at level $n=0$, which is defined by $s_*=\mu(0)$.  Assuming the spectral function does not have any double zeros, it exhibits a single root labeled by $\nu_{*}(s_*) = 0$, whose vanishing corresponds to the factorization channel.  To take the residue, we change variables from $s_*$ to $\nu_{*}(s_*)$ at the expense of a Jacobian, $\lim\limits_{s_*\rightarrow \mu(0)}(s_*-\mu(0))/\nu_{*}(s_*)=\mu'(0)$, yielding
\eq{
\lim_{s_*\rightarrow \mu(0) }(s_*-\mu(0)) A(\{ s_I\}) &= \mu'(0)  \lim_{\nu_{*}(s_*)\rightarrow 0} \left( \prod_{I} \sum_{\alpha_{I}} \right)\nu_{*}(s_*) \AV(\{ \nu_{\alpha_I}(s_I)\})\\
&= \mu'(0)  \left( \prod_{I\neq*} \sum_{\alpha_{I}} \right) \lim_{\nu_{*}(s_*)\rightarrow 0}\nu_{*}(s_*) 
\AV(\{  \nu_{\alpha_{I}}(s_{I})\}) |_{\alpha_*=*}\\
&= \mu'(0) \left( \prod_{I\neq 0} \sum_{\alpha_{I}} \right)  \sum_{L \perp R} \AV( \{\nu_{\alpha_{I_L}}(s_{I_L})\})\AV(\{\nu_{\alpha_{I_R}}(s_{I_R})\}).
}{eq:factorpre}
Note that in second line, we have dropped the sum over $\alpha_*$, which labels the roots folding the factorizing invariant $s_*$.  The only term from this sum that is included is the one in which $s_*$ is folded into $\nu_*(s_*)$.  For all the other terms, the resulting contribution will not have a singularity at $s_*=\mu(0)$, so they vanish when taking the residue.  In the third line, we have simply used the factorization condition on the Veneziano amplitude in \Eq{eq:factorV}.

At this point, it is useful to consider the counting of kinematic invariants.  The total number of external legs is partitioned according to $N=|L|+|R|$.
The left-hand side of  \Eq{eq:factorpre}  is the factorization limit of a higher-point amplitude with $N$ external particles, so it depends on $N(N\,{-}\,3)/2 - 1$ kinematic invariants.  Meanwhile, the right-hand side has lower-point amplitudes with $|L|+1$ and $|R|+1$ external particles, corresponding to $(|L|+1)(|L|-2)/2$ and $(|R|+1)(|R|-2)/2$ invariants.
The difference between these numbers is $(|L|-1)(|R|-1)$, which corresponds to the number of Galois sums in \Eq{eq:factorpre} that act trivially, since there is no dependence on their corresponding kinematic invariants on the factorization channel.  Since these sums run over $h$ roots, they introduce a trivial multiplicity factor of $h^{(|L|-1)(|R|-1)}$. By splitting this factor between the lower-point amplitudes, we obtain
\be
\begin{aligned}
&\lim_{s_*\rightarrow \mu(0) }(s_*-\mu(0)) A(\{ s_I\})  \\& = \mu'(0)\! \sum_{L \perp R} \left[h^{(|L|-1)}\left( \prod_{I_L} \sum_{\alpha_{I_L}} \right) \AV( \{\nu_{\alpha_{I_L}}(s_{I_L})\})\right]\left[h^{(|R|-1)}\left( \prod_{I_R} \sum_{\alpha_{I_R}} \right)\AV(\{\nu_{\alpha_{I_R}}(s_{I_R})\})\right]\! ,
\end{aligned}
\ee
where we have rearranged the original sum over $I$ into sums over $I_L$ and $I_R$.
We immediately recognize the objects in brackets as lower-point bespoke amplitudes, so
\be
\lim_{s_*\rightarrow \mu(0) }(s_*-\mu(0)) A(\{ s_I\})= \mu'(0) \sum_{L \perp R} h^{(|L|-1)}A( \{\nu_{\alpha_{I_L}}(s_{I_L})\}) h^{(|R|-1)}A(\{\nu_{\alpha_{I_R}}(s_{I_R})\}).
\ee
Thus, our bespoke amplitudes have factorized from higher-point to lower-point, modulo a multiplicative constant that depends on the numbers of external particles.  While the latter may seem peculiar, we emphasize that these factors can be trivially eliminated.  For example, $\mu'(0)$ can be reabsorbed into the normalization of the $n=0$ state.  Meanwhile, the factor of $h^{(|L|-1)(|R|-1)}$ will not arise if each Galois sum in \Eq{eq:Mn} is replaced with a Galois {\it average}, by which we mean the Galois sum divided by the multiplicity of roots $h$.

\section{Discussion}

In this paper, we have derived analytic expressions for a new family of scattering amplitudes that simultaneously exhibit a litany of remarkable properties: 
\begin{enumerate}[label=(\alph*)]
\item {\bf Arbitrary Spectrum:} The spectrum of masses is described by a rational polynomial of arbitrary degree.  This function can be specified to interpolate through any finite collection of masses and exhibit any type of polynomial growth.\label{item:spectrum}
\item {\bf Infinite Spin Tower:} The spectrum describes states of arbitrarily high spin.  However, to maintain locality, only a finite number of spins are exchanged at each resonance.
\item {\bf Tame Ultraviolet Behavior:} The high-energy Regge behavior, $s\rightarrow \infty$, is not just polynomial bounded, but either vanishing or constant at some fixed $t$.
\item  {\bf Dual Resonance:}  The amplitude has a closed-form representation in terms of an infinite sum over purely $s$-channel or $t$-channel exchanges.
\item {\bf Meromorphicity:} The amplitude is analytic but for simple poles.  In particular, there are no branch-cut singularities in the kinematic variables.\label{item:analytic}
\item {\bf Integral Representations:}  The amplitude can be represented as an integral over moduli, strongly reminiscent of a worldsheet construction.  As a bonus, the amplitude can also be expressed as an elegant $\dlog$ transform of existing string amplitudes.\label{item:integral}
\item {\bf Higher-Point Generalization:} Our construction has a natural generalization to higher-point, which produces candidate amplitudes with the above properties.\label{item:higherpoint} 
\end{enumerate}
The scattering amplitudes of string theory are the {\it only} objects hitherto known to possess {\it almost} all of these properties, of course with the exception of \ref{item:spectrum}. 
The recently constructed class of hypergeometric amplitudes~\cite{CheungRemmen2023} does not yet have a generalization to \ref{item:higherpoint} and, with a spectrum matching string theory, does not exhibit property \ref{item:spectrum}.
Meanwhile, the Coon amplitude~\cite{Coon} and its generalizations~\cite{CheungRemmen2022,CheungRemmen2023} do not satisfy conditions \ref{item:spectrum}, \ref{item:analytic}, or \ref{item:integral}.

That the properties above can all be satisfied at once is both surprising and compelling.
Fundamentally, this miracle is possible because our bespoke amplitudes are directly constructed from a Galois sum---or equivalently, a $\dlog$ integral transform---of the Veneziano amplitude.  For this reason, the bespoke amplitudes inherit many of the wondrous properties of the string, albeit without a stringy spectrum.   The constraints of partial wave unitarity are substantial, but they leave open a broad region of putatively consistent amplitudes.

The possible avenues for future work are numerous.  Most pressingly, there is the critical question of whether our bespoke amplitudes possess a bona fide, underlying physical description.  While this is far from guaranteed---and the space of amplitudes we have constructed is not a {\it theory}---the integral representations presented in Sec.~\ref{sec:worldsheet} offer enticing hope.  That said, it is clear that if such a theory exists, it will likely require some unusual dynamics.  

Furthermore, it would be interesting to pursue other, broader applications of our construction.   For example, while the present paper utilizes the Veneziano amplitude of the bosonic string as its basic building block, it would be straightforward to instead use the amplitudes of the superstring.  Such an approach would yield dual resonant amplitudes that reduce to Yang-Mills theory at low energies but support a customizable spectrum. 
Another question of more phenomenological interest would be to apply these ideas to the strong interactions.  Perhaps one can construct a dual resonant scattering amplitude that interpolates through the spectrum of mesons!

Last but not least is the open question of gravity.  In particular, one can ask  whether there is an analogue of our construction that is applicable to the amplitudes of the closed string.  
Here an immediate difficulty arises from the fact that graviton scattering exhibits full crossing symmetry among $s$, $t$, and $u$.  The on-shell condition, $s+t+u=0$, cannot be preserved under any nonlinear, permutation invariant remapping the kinematic variables, since $\nu(s)+\nu(t)+\nu(u)\neq0$.  This basic problem may be a hint that gravitational amplitudes are inherently less amenable to modification than their gauge theory cousins.

\vspace{1em}

\begin{center} 
{\bf Acknowledgments}
\end{center}
\noindent 
We thank Nathaniel Craig, Nick Geiser, David Gross, Aaron Hillman, Igor Klebanov, Piotr Tourkine, and Sasha Zhiboedov for comments.
 C.C. is supported by the Department of Energy (Grant No.~DE-SC0011632) and by the Walter Burke Institute for Theoretical Physics.
G.N.R. is supported by the James Arthur Postdoctoral Fellowship at New York University, and was supported at the Kavli Institute for Theoretical Physics by the Simons Foundation
(Grant No.~216179) and the National Science Foundation (Grant No.~NSF PHY-1748958)
and at the University of California, Santa Barbara by the Fundamental Physics Fellowship.

\vspace{1em}

\appendix 

\section{Partial Wave Unitarity}\label{app:unitarity}

For completeness, let us review here the formalism of partial wave unitarity.  We decompose
the residue of an amplitude at level $n$ in terms of partial waves,
\be
R(n,t) = \sum_\ell a_{n,\ell} G_\ell^{(D)}(x), 
\ee
where  $x=\cos\theta$ is related to the scattering angle and $G_\ell^{(D)}$ is the Gegenbauer polynomial in $D\geq 4$ spacetime dimensions.
In a unitary theory, the coefficients of the partial waves on each pole must have all positive coefficients, so $a_{n,\ell} \geq 0$.
On the residue, $t$ is related to the scattering angle in terms of $x$ via
\eq{
t =\frac12(x-1) (\mu(n)-4m_{\rm ext}^2),
}{}
 where $m_{\rm ext}^2$ gives the mass squared of the four external particles, which we take to be equal. 
In order to compute $a_{n,\ell}$ it will be convenient to use the monomial identity,
\eq{
x^k = \sum_{j=0}^{\lfloor{k/2\rfloor}}  \frac{k!\left[1+\frac{2(k-2j)}{D-3}\right]\Gamma\left(\frac{D-1}{2}\right)}{2^k j!\Gamma\left(\frac{D-1}{2} + k-j\right)} G_{k-2j}^{(D)}(x).
}{xk}
By expressing the residue as $R(n,t(x))$ and replacing each power in $x$ via \Eq{xk}, we can directly read off the partial wave coefficients $a_{n,\ell}$.

An amusing mathematical fact is that the Gegenbauer polynomials in $D$ dimensions are always expressible as nonnegative sums of the Gegenbauer polynomials in $D'\leq D$ dimensions.
In other words, the connection coefficients associated with the Gegenbauer polynomials in different dimensions are nonnegative~\cite{Maroni}.
This observation has the profound physical implication that unitarity is {\it preserved under dimensional reduction}.
On the other hand, an amplitude may cease to be unitary if the dimensionality of the spacetime is increased.  For the Veneziano amplitude, this fact is a tree-level manifestation of the critical dimension of string theory~\cite{Arkani-Hamed:2022gsa}.
One can indeed verify that, for example, with $m_{\rm ext}^2 = 0$, the residues of the Veneziano amplitude in \Eq{eq:RV} are nonnegative in $D \leq 10$, while for $m_{\rm ext}^2 = -1$ they are nonnegative for $D\leq 26$, corresponding to the superstring and bosonic string, respectively.

On the other hand, it appears that there are always negative residues in the Kaluza-Klein amplitude defined in \Eq{eq:Rtoy}.   Concretely, we find that for arbitrary $\delta$, $m_{\rm ext}^2$, and $D\geq 4$, there exists some partial wave for which $a_{n,\ell}<0$ for $n\leq 3$.
This is consistent with the no-go result of Ref.~\cite{Zohar}.
In particular, Ref.~\cite{Zohar} considered a class of crossing-symmetric, meromorphic amplitudes satisfying i)~partial wave unitarity, ii)~Regge boundedness, iii)~Regge behavior in the eikonal hard scattering limit (small fixed $t/s$ with $s,t\rightarrow\infty$), and iv)~a spectrum free of accumulation points.
For such objects, they showed that the Veneziano amplitude is asymptotically unique: $\mu(n) \sim n$ at large $n$, and the hard scattering limit takes the form given in Eq.~\eqref{eq:AVhard}.
However, such a conclusion hinges on an additional technical assumption made in Ref.~\cite{Zohar}, namely that certain  ``excess zeros'' in the $s$-plane at fixed $t$ must lie in a particular ellipse.
While some arguments and numerical evidence for this assumption were given in Ref.~\cite{Zohar}, a rigorous proof was not provided, and new results~\cite{SashaTalk} have indicated that it can be violated.
We therefore take the asymptotic uniqueness result  as simply motivation for our requirement that $P$ and $Q$ differ in degree by one, but this stipulation is unimportant for our formalism.
As we have seen, our bespoke construction straightforwardly allows for dual resonant amplitudes with spectra of arbitrary degree. 
As always, however, one must numerically check partial wave unitarity ex post facto.

\vspace{\baselineskip}

\bibliographystyle{utphys-modified}
\bibliography{Bespoke_Dual_Resonance}

\end{document}